\documentclass{aa}

\usepackage{graphicx,natbib}
\newcommand{\pun}[1]{\mbox{\rm\,#1}} 
\newcommand{\punms}{\mbox{\rm\,m\,s$^{-1}$}}
\newcommand{\punkms}{\mbox{\rm\,km\,s$^{-1}$}}

\newcommand{\logg}{\ensuremath{\log g}}

\newcommand{\mlp}{\ensuremath{\alpha_{\mathrm{MLT}}}}

\newcommand{\Teff}{\ensuremath{T_{\mathrm{eff}}}}
\newcommand{\tauross}{\ensuremath{\tau_{\mathrm{ross}}}}

\newcommand{\draftflag}{false}

\newcommand{\beq}{\begin{equation}}
\newcommand{\eeq}{\end{equation}}

\newcommand{\sig}[1]{{\ensuremath{\sigma_{#1}}}}
\newcommand{\cov}[2]{{\ensuremath{\mathrm{C}\left[#1,#2\right]}}}
\newcommand{\xtmean}[1]{\ensuremath{\left\langle #1\right\rangle}}

\newcommand{\eref}[1]{\mbox{(\ref{#1})}}

\newcommand{\cp}{\ensuremath{c_{\mathrm{p}}}}
\newcommand{\taueddy}{\ensuremath{\tau_{\mathrm{e}}}}

\newcommand{\lmix}{\ensuremath{\Lambda}}
\newcommand{\Hp}{\ensuremath{H_{\mathrm{P}}}}

\newcommand{\moch}{C5}
\newcommand{\mocm}{C4}
\newcommand{\mocl}{C3}
\newcommand{\mohm}{H4}
\newcommand{\mohx}{HX}
\newcommand{\mosu}{S}
\newcommand{\mosg}{SG}

\newcommand{\fex}{\ensuremath{f_\mathrm{ex}}}
\newcommand{\Hfex}{\ensuremath{H_\mathrm{fex}}}
\newcommand{\Fmup}{\ensuremath{F_\mathrm{mass}^\mathrm{up}}}
\newcommand{\mcol}{\ensuremath{m_\mathrm{col}}}
\newcommand{\lmode}{\ensuremath{\lambda_\mathrm{mode}}}
\newcommand{\Hpsurf}{\ensuremath{H_\mathrm{P}^\mathrm{surf}}}
\newcommand{\Psurf}{\ensuremath{P_\mathrm{surf}}}

\newcommand{\vmix}{\ensuremath{v_\mathrm{mix}}}
\newcommand{\dirms}{\ensuremath{\delta I_\mathrm{rms}/I}}
\newcommand{\Pe}{\ensuremath{\mathrm{Pe}}}
\newcommand{\blank}{\hspace{1ex}}

\newcommand{\changed}[1]{#1}

\begin{document}

\title{Energy transport, overshoot, and mixing in the atmospheres of
M-type main- and pre-main-sequence objects}
\titlerunning{Radiation-hydrodynamics of M-type atmospheres}
\authorrunning{Ludwig, Allard, Hauschildt}
\offprints{Hans-G\"unter Ludwig}

\author{ Hans-G\"unter Ludwig\inst{1,2,3} \and France Allard\inst{3,4}
  \and Peter H. Hauschildt\inst{5,3}}

\institute{GEPI, CIFIST, Observatoire de Paris-Meudon, 5 place Jules Janssen, 92195 Meudon
  Cedex, France\\ \email{Hans.Ludwig@obspm.fr} 
  \and
  Lund Observatory, Lund University, Box~43, 22100 Lund, Sweden
  \and 
  Ecole Normale Sup{\'e}rieure de Lyon, Centre de Recherche Astronomique de  
  Lyon, 46 all{\'e}e d'Italie, 69364 Lyon Cedex 07, France; CNRS, UMR 5574;
  Universit{\'e} de Lyon~1, Lyon, France
  \and
  Institut d'Astrophysique de Paris, CNRS, UMR 7095,
  98$^{\mathrm{bis}}$ boulevard Arago, 75014 Paris, France;
  Universit{\'e} Pierre et Marie Curie-Paris 6, 75005 Paris, France
  \and
  Hamburger Sternwarte, Gojenbergsweg~112, 21029 Hamburg, Germany
}

\date{Received date; accepted date}

\abstract{We constructed hydrodynamical model atmospheres for mid M-type
  main-, as well as pre-main-sequence (PMS) objects. Despite the complex
  chemistry encountered in these cool atmospheres a reasonably accurate
  representation of the radiative transfer is possible, even in the context of
  time-dependent and three-dimensional models. The models provide detailed
  information about the morphology of M-type granulation and statistical
  properties of the convective surface flows.  In particular, we determined
  the efficiency of the convective energy transport, and the efficiency of
  mixing by convective overshoot. The convective transport efficiency was
  expressed in terms of an equivalent mixing-length parameter \mlp\ in the
  formulation of mixing-length theory (MLT) given by \citet{Mihalas78}. \mlp\
  amounts to values around $\approx 2$ for matching the entropy of the deep,
  adiabatically stratified regions of the convective envelope, and lies
  between 2.5 and 3.0 for matching the thermal structure of the deep
  photosphere.  For current spectral analysis of PMS objects this implies that
  MLT models based on $\mlp=2.0$ overestimate the effective temperature by
  100\pun{K} and surface gravities by 0.25\pun{dex}. The average thermal
  structure of the formally convectively stable layers is little affected by
  convective overshoot and wave heating, i.e., stays close to radiative
  equilibrium conditions. Our models suggest that the rate of mixing by
  convective overshoot declines exponentially with geometrical distance to the
  Schwarzschild stability boundary. \changed{It increases at given effective
  temperature with decreasing gravitational acceleration.}
} 

\maketitle

\section{Introduction}

The increasing number of stars, brown dwarfs, and extrasolar planets of
spectral class M or later discovered by infrared surveys and radial velocity
searches has spawned a great deal of interest in the atmospheric physics of
these objects.  Their atmospheres are substantially cooler than the solar
atmosphere, allowing the formation of molecules, or even liquid and solid
condensates. Convection is a ubiquitous phenomenon in these atmospheres
shaping their thermal structure and the distribution of chemical species.
Hydrodynamical simulations of solar and stellar granulation including a
realistic description of radiative transfer have become an increasingly
powerful and handy instrument for studying the influence of convective flows
on the structure of late-type stellar atmospheres as well as on the
formation of their spectra
\citep[e.g.,][]{Nordlund82,Steffen+al89,Chan+Sofia89,Nordlund+Dravins90,CBTMH91,Ludwig+al94,Gadun+Pikalov96,
  Steiner+al98,Stein+Nordlund98,Asplund+al99b,Voegler+Schuessler03,Robinson+al04}.
Here we report on efforts to construct hydrodynamical model atmospheres for
mid M-type objects. The spectral type just borders the temperature where the
formation of condensates becomes important.  The motivation of this
investigation was twofold: first, pre-main-sequence (PMS) evolutionary
models of M-type stars and brown dwarfs based on mixing-length theory
\citep[MLT,][]{BoehmVitense58} to describe the convective energy transport
depend sensitively on the poorly constrained mixing-length
parameter\footnote{The ratio between the mixing-length and local pressure
  scale height.}~\mlp\ \citep{Baraffe+al02}. Our hydrodynamical models
represent convection essentially from first principles, and are free of the
uncertainties of MLT allowing to put the stellar models on a firmer footing.
Second, the distribution of dust clouds in cool brown dwarfs depends on the
efficiency of mixing of their atmospheres by convective overshoot
\citep{Ackerman+Marley01,Allard+al03,Helling+al04}. For other work on the
modeling of dust cloud formation in very low mass stars and brown dwarfs, see e.g.
\citet{Cooper+al03} and \citet{Tsuji05}, and references therein. While by
construction MLT cannot describe convective overshoot it is naturally
represented in our three-dimensional hydrodynamical models.

The present investigation is extending a previous study of an M-dwarf
atmosphere by \citet[][hereafter LAH]{Ludwig+al02} to PMS objects at lower
surface gravity.  A preliminary account of the results was given in
\citet{Ludwig03}. We start in Sect.~\ref{s:modelover} with an overview of the
model construction, in particular related to approximations we adopted in the
radiative transfer.  In Sect.~\ref{s:morphology} we discuss the general
morphology of the convective flows in the M-type atmospheres and present
some statistical properties. In Sect.~\ref{s:convection} we provide estimates of
the efficiency of the convective energy transport in terms of an effective
mixing-length parameter, and discuss consequences for the analysis of PMS
M-type objects in the framework of present standard model atmospheres.  We
continue in Sect.~\ref{s:mixing} by characterizing the properties of the
atmospheric mixing found in the hydrodynamical models, and conclude with
final remarks in Sect.~\ref{s:remarks}. In our investigation we take repeatedly
recourse to the solar atmosphere as standard benchmark. 

\section{Model overview}
\label{s:modelover}

\begin{table*}
\begin{flushleft}
\caption[]{%
  The RHD models discussed in the paper: ID is the identifier used to refer to
  a model in this paper, \Teff\ the effective temperature of the model
  including an estimate of its RMS fluctuations, \logg\ the preset
  gravitational acceleration, Size is the geometrical size of the
  computational domain (XxYxZ, where Z denotes the vertical, X and Y the
  horizontal directions), \Hpsurf\ the pressure scale height at the surface,
  \Psurf\ the pressure at the surface, $v_\mathrm{rms}^\mathrm{max}$ the
  maximum vertical RMS velocity in the convective layers, \dirms\ the relative
  intensity contrast, \mlp\,(evo) the mixing-length necessary to match the
  asymptotic entropy (see Sect.~\ref{s:convection}), \mlp\,(phot) the
  mixing-length parameter necessary to match the model's temperature in the
  deep photosphere, \Hfex\ the scale height of the decline of the atmospheric
  mixing rate (see Sect.~\ref{s:mixing}), and Modelcode an internal model
  identifier. Parenthesis indicate uncertain values.}
\label{t:mhdmodels}
\begin{tabular}{lllllllllllll}
\hline\noalign{\smallskip}
ID    & \Teff          & \logg& Size               &\Hpsurf&\Psurf    & $v_\mathrm{rms}^\mathrm{max}$& \dirms& \mlp  & \mlp    & \Hfex & Modelcode \\
      & [K]            &      & [Mm]               &[Mm]   &$\log_{10}$& [\punms]                    &[\%]   & (evo) & (phot) & [\Hp] &           \\
\noalign{\smallskip}
\hline\noalign{\smallskip}
\moch & $2789 \pm 0.7$ & 5.0  & 0.25x0.25x0.087    & 0.012 & 6.1 & 240 & 1.2 & (1.5) & (2.5) & 0.5 & d3gt30g50n18\\
\mocm & $2800 \pm 1$   & 4.0  & 3.75x3.75x1.16     & 0.13  & 5.3 & 450 & 3.0 &  2.1  &  2.5  & 3.2 & d3gt30g40n1\\
\mocl & $2800 \pm 2.7$ & 3.0  & 37.5x37.5x13.8     & 1.5   & 4.5 & 820 & 8.2 &  2.1  &  2.8  & (18)& d3gt30g30n1\\
\mohm & $3280 \pm 2.8$ & 4.0  & 4.38x4.38x1.54     & 0.19  & 5.2 & 690 & 5.4 &  1.85 &  3.0  & (28)& d3gt33g40n1\\
\mohx & $3275 \pm 2.8$ & 4.0  & 4.38x4.38x1.81     & 0.19  & 5.2 & 690 & 5.6 &  -    &  -    & -   & d3gt33g40n2\\
\mosu & $5640 \pm 14$  & 4.44 & 6.0x6.0x3.2        & 0.15  & 5.1 & 2600 & 16  & -    &  -    & 2.4 & sun3d      \\
\mosg & $4610 \pm 23$  & 2.94 & 141x141x95.3       & 3.8   & 4.4 & 3400 & 21  & -    &  -    & -   & d3gt45g29n1\\
\noalign{\smallskip}
\hline
\end{tabular}%
\end{flushleft}
\end{table*}

\begin{figure}
\resizebox{\hsize}{!}{\includegraphics[draft = \draftflag]%
{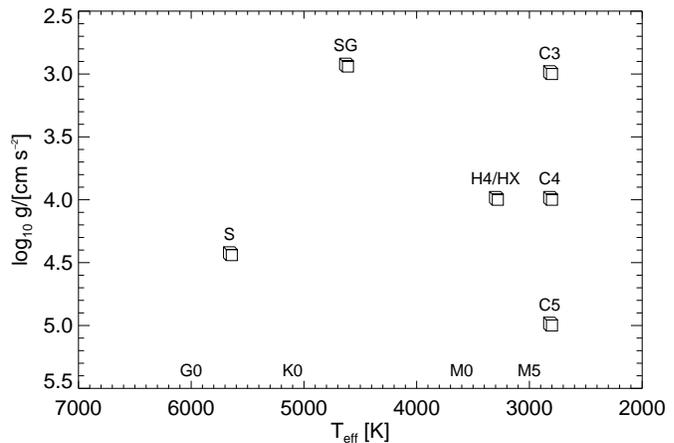}}
\caption[]{The radiation-hydrodynamics models in the effective
temperature-gravity plane (cubes). The models are labeled by their IDs used
for reference in this paper. The approximate spectral class is
indicated at the temperature axis.}
\label{f:modover}
\end{figure}

\changed{%
Figure~\ref{f:modover} illustrates the positions of our hydrodynamical model
atmospheres in the \Teff-\logg-plane: three M-type models are located close to
$\Teff=2800\pun{K}$ with \logg=3.0, 4.0, and 5.0 which form a \logg-sequence.
In the following we shall refer to them as models~\mocl, \mocm, and \moch,
respectively.  To assess temperature effects, a 500\pun{K} hotter M-type model
was constructed at \Teff=3280\pun{K} and \logg=4.0.  For further comparison we
also considered a solar (in terms of its hydrodynamical properties)
model~\mosu\ at \Teff=5640\pun{K}, \logg=4.44, and a model~\mosg\ of a
subgiant at 4610\pun{K} and \logg=2.94.

For investigating the influence of the position of the upper boundary
condition we constructed an additional model~\mohx\ with the same atmospheric
parameters as model~\mohm\ but extending 270\pun{km} (corresponding to
3.7\pun{\Hp}) higher up than model~\mohm. The flow in the extended region
exhibits larger fluctuations than encountered in deeper layers favoring the
formation of sharp flow features.  For reasons of numerical stability we had
to increase the numerical viscosity so that the model is not fully
differentially comparable to model~\mohm.  Nevertheless, it should give an
indication of the level of the influence of the upper boundary, in particular
when the upper boundary is located close to the convectively unstable region.
In plots\footnote{We included data of model~\mohx\ in Figs.~\ref{f:temps},
  \ref{f:vrms}, \ref{f:pturb}, \ref{f:dtrms}, \ref{f:dprms}, \ref{f:entropy},
  \ref{f:temps2}, and~\ref{f:velmlt}} that follow, we depict model~\mohx\ 
always as triple-dot-dashed line without labeling it by its name like the
other models.  Naturally, the behavior of model~\mohx\ closely follows
model~\mohm\ in the deeper layers so that its connection to model~\mohm\ is
readily apparent.

All models were evolved until a thermally and dynamically relaxed state was
reached. All models except~\mohx\ have 125x125x82 grid points (XxYxZ
direction), \mohx\ has 125x125x102 grid points due its larger vertical extent.
The numerical grid is equidistant in x- and y- direction while in vertical
z-direction the grid spacing is to first order chosen to provide the same
number of grid points per pressure scale height. In addition, the resolution
is increased in layers around continuum optical depth unity if a steep
vertical temperature gradient is present.  All models have solar chemical
composition.  Table~\ref{t:mhdmodels} summarizes their properties.
} 

The overall methodology applied in this work is the same as in LAH, and we
refer the reader to this paper for details beyond the short description we
provide here.  

The radiation-hydrodynamics (RHD) simulations were performed with a convection
code developed by {\AA}.~ Nordlund and. R.F.~Stein \citep[see][and references
therein]{Stein+Nordlund98}.  The code solves the hydrodynamical equations of
compressible gas dynamics coupled with non-local radiative transfer in three
spatial dimensions. The time-independent radiative transfer is treated
assuming strict LTE.  The wavelength dependence of the radiation field is
represented by a small number of wavelength bins. Open lower and upper
boundaries, as well as periodic lateral boundaries are assumed. The effective
temperature of a model (i.e., the average emergent radiative flux) is
controlled indirectly by prescribing the entropy of inflowing material at the
lower boundary. Magnetic fields and rotation are neglected.  Opacities and
equation-of-state have been adapted to the conditions encountered in
M-type atmospheres. \changed{The equation of state includes the ionization of
  H and He , as well as
$\mathrm{H}_2$ molecule formation according to Saha-Boltzmann statistics.
$\mathrm{H}_2$ molecule formation is the thermodynamically most important
process in the M-type atmospheres. The opacities include contributions of
molecular lines but neglect contributions of dust grains which is a good
approximation at the temperatures prevailing in the models.} 
The opacities were extracted from the opacity data base of the PHOENIX model
atmosphere code \citep[for a description of PHOENIX and corresponding
opacities see][]{Hauschildt+al99,Ferguson+al05}.

\subsection{Radiative transfer}

\begin{figure}[!tb]
\resizebox{\hsize}{!}{\includegraphics[draft = \draftflag]%
{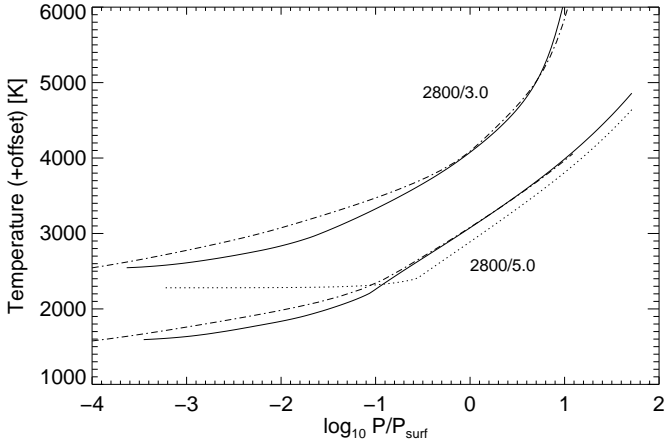}}
\caption[]{Comparison of the pressure-temperature structure between 1D
hydrostatic model atmospheres in radiative-convective equilibrium
based on the OBM (solid lines) with four wavelength groups, and PHOENIX
models based on direct opacity sampling employing several $10^4$
wavelength points (dash-dotted). Shown are two examples at
\Teff=2800\,K, \logg=3.0 and 5.0, respectively. For clarity, the
\logg=3.0 models were shifted by +1000\,K. Also shown is a model
employing grey radiative transfer (dotted line). The pressure is given
in units of the pressure at Rosseland optical depth
unity~$P_{\mathrm{surf}}$.}
\label{f:obmmdwarfs}
\end{figure}

We want to derive quantitative estimates of the mixing by convective
overshoot, as well as obtain a measure of the efficiency of the convective
energy transport. For addressing these issues, the RHD models have to give a
reasonably accurate representation of the actual atmospheric conditions. Here
we are particularly concerned about the radiative energy transport, which is
complicated by the huge number of molecular absorption lines. In our RHD
models, we use a multigroup technique (dubbed {\it Opacity Binning Method\/},
hereafter OBM) for modeling the radiative energy exchange which employs four
groups for representing the wavelength dependence of the radiation field
\citep[][LAH]{Nordlund82,Ludwig92,Ludwig+al94,Voegler+al04}. The wavelength
groups have been optimized for an atmosphere at \Teff=2900\,K and \logg=5.3.
Figure~\ref{f:obmmdwarfs} illustrates the accuracy which is achieved with the
OBM for the present models. We compare 1D MLT model structures (\mlp=1.0) in
radiative-convective equilibrium computed with the OBM approximation and
high-precision opacity sampling.  While there are differences between the
atmospheric structures, the OBM nevertheless provides a significant
improvement with respect to a simple grey approximation.  Temperature
differences get larger as one moves away from the atmospheric parameters the
OBM was optimized for, and reach up to 250\pun{K} in the model at
2800\pun{K} and \logg=3.0. However, in the present context it is not so much
the absolute temperature error as the change of the temperature gradient which
is relevant.

\changed{%
Convection is driven by buoyancy forces whose dynamical effects scale with
  the entropy gradient. The thermal structure is controlled by a balance
  between radiative and advective (due to compression or expansion of mass
  elements) heating (or cooling) of which the P{\'e}clet number (see
  appendix~\ref{s:peclet} for its definition and computation) is a convenient
  dimensionless measure. In the deeper photospheric layers, the OBM profiles
  shown in Fig.~\ref{f:obmmdwarfs} have steeper temperature gradients than the
  profiles based on opacity sampling. For a given velocity, this makes the
  rate of temperature change of a mass element moving in vertical direction
  larger.  In other words, the time scale of advection related temperature
  changes becomes shorter, and the P{\'e}clet number larger as long as the
  radiative time scales remain the same. The opposite behavior is present in
  the higher photospheric layers.  The typical P{\'e}clet numbers turn out to
  be by a factor of 1.2 times larger for the \logg=5.0, and by up to a factor
  of 2 times smaller for the \logg=3.0 model comparing OBM to the opacity
  sampling stratifications. 
}

To mitigate the effects of the shortcomings, primarily related to the OBM
approximation in the radiative transfer, we took a {\em differential
  approach\/} when measuring model properties: whenever possible we compared
RHD and hydrostatic model atmospheres based on the same OBM radiative transfer
scheme since we were interested to study the systematic change of model
properties with effective temperature and gravitational acceleration. However,
there remains the possibility that even this differential approach cannot
fully eliminate the impact of the artificial shift of the P{\'e}clet number
introduced by the OBM approximation among the models. This has to be kept in
mind when interpreting results later.

\section{Morphology of granulation in M-type objects}
\label{s:morphology}

\begin{figure*}[!t]
\resizebox{\hsize}{!}{\includegraphics[draft = \draftflag]%
  {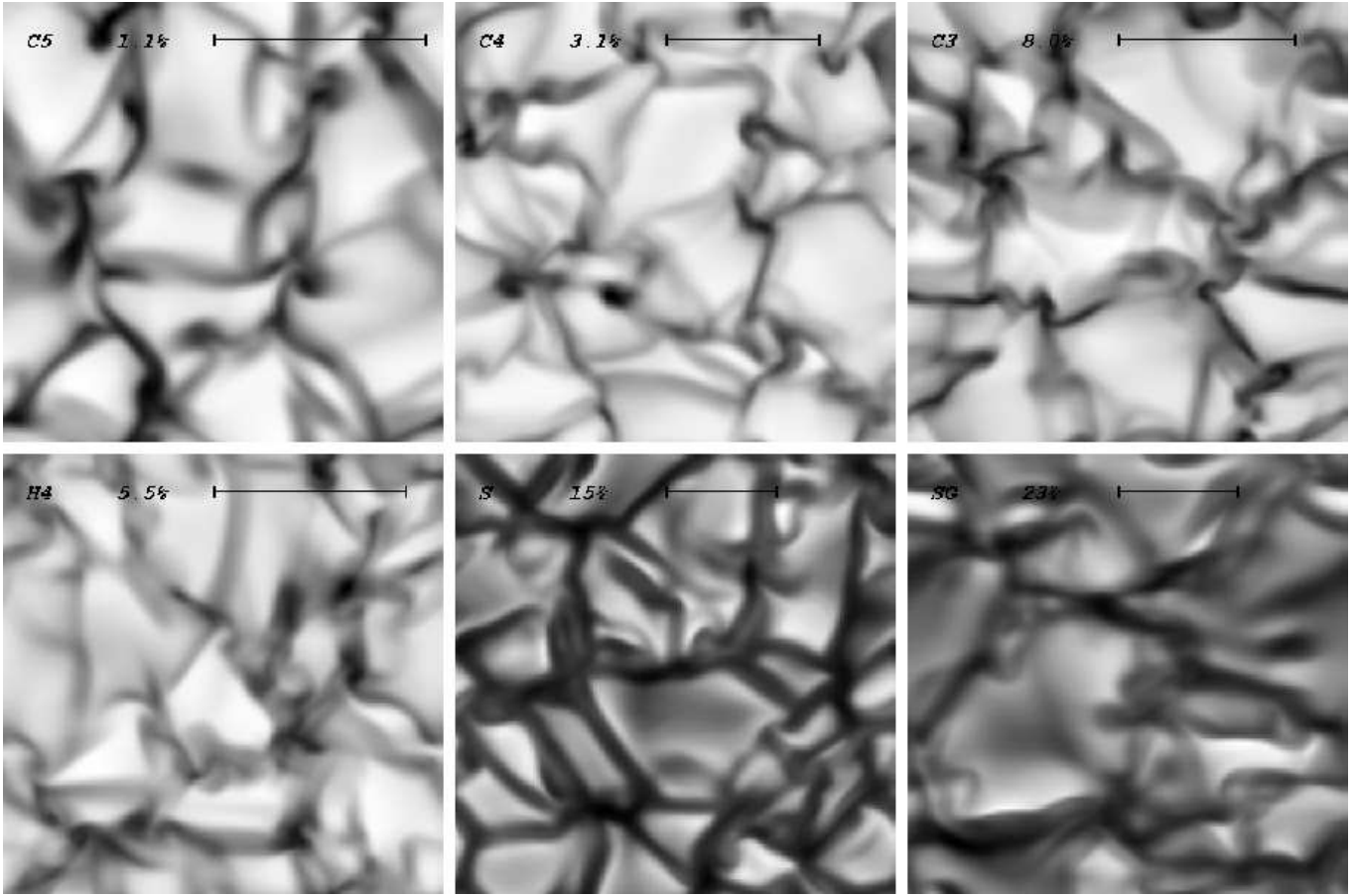}}
\caption[]{Granulation patterns of the six RHD models. The model IDs are
  given in the upper left corners (see Tab.~\ref{t:mhdmodels}) of the
  images. Shown are snapshots of the emergent intensity in the
  continuum. For each image a separate grey scale is used, lighter
  shades correspond to higher intensities. The relative intensity
  contrast of the particular image is stated, and the  bar indicates a
  length of $10\times\Hpsurf$.}
\label{f:grancomp}
\end{figure*}

Figure~\ref{f:grancomp} shows an inter-comparison of the granulation patterns
typically encountered during the temporal evolution of our RHD models.  The
first thing to note is that surface convection in M-type objects produces a
granular pattern qualitatively resembling solar-type granulation: bright
extended regions of up-welling material which are surrounded by dark
concentrated lanes of down-flowing material. The dark lanes form an
interconnected network.  Looking more closely, however, granules are less
regularly delineated in M-type objects. The inter-granular lanes show a higher
degree of variability in terms of their strength -- in particular in
comparison to the solar model~\mosu, to lesser extend in comparison to the
subgiant model~\mosg. A feature which is uncommon in the solar granulation pattern
are the dark ``knots'' found in or attached to the inter-granular lanes
prominent in the M-type models \moch\ and \mocm. The knots are associated with
strong downdrafts which carry a significant vertical component of angular
momentum.

\changed{We have no convincing explanation at hand why convective flows in
  M-type atmospheres tend to form such vortical structures. It is unlikely
  that they are (as suspected by the referee) an artifact related to the
  boundary condition since the structures do not reach up into layers close to
  it. Moreover, the test model~\mohx\ shows also the knots despite the fact
  that its upper boundary is located about 10\pun{\Hp} above the continuum
  forming layers.  The presence or absence of knots may rather be related to
  the level of horizontal shearing in the layers around optical depth unity
  where convective driving is usually strongest and is important for the
  formation of flow structures.  Figure~\ref{f:vrms} illustrates that under
  solar-like conditions (models~\mosu\ and~\mosg) shear flows are common in
  the surface layers, and are much more pronounced than in the M-type
  atmospheres. Such shear flows make the formation of structures extending in
  vertical direction difficult, and may be the reason why the knots do not
  appear in the solar-like models and are less developed in models~\mohm\ 
  and~\mocl.} 

\changed{As a side-point we would like to remark that it is not immediately
  clear why the knots appear in fact dark. However, centrifugal forces in the
  vortices are evidently not strong enough to lead to an evacuation of their
  interior on a level that would allow radiation from deeper, hotter layers to
  escape and let them appear brighter than their surroundings. The conditions
  are not as extreme as for bright flux tubes in the solar photosphere where
  magnetic pressure allows for a substantial degree of evacuation.}

\begin{figure}
\resizebox{\hsize}{!}{\includegraphics[draft = \draftflag]%
{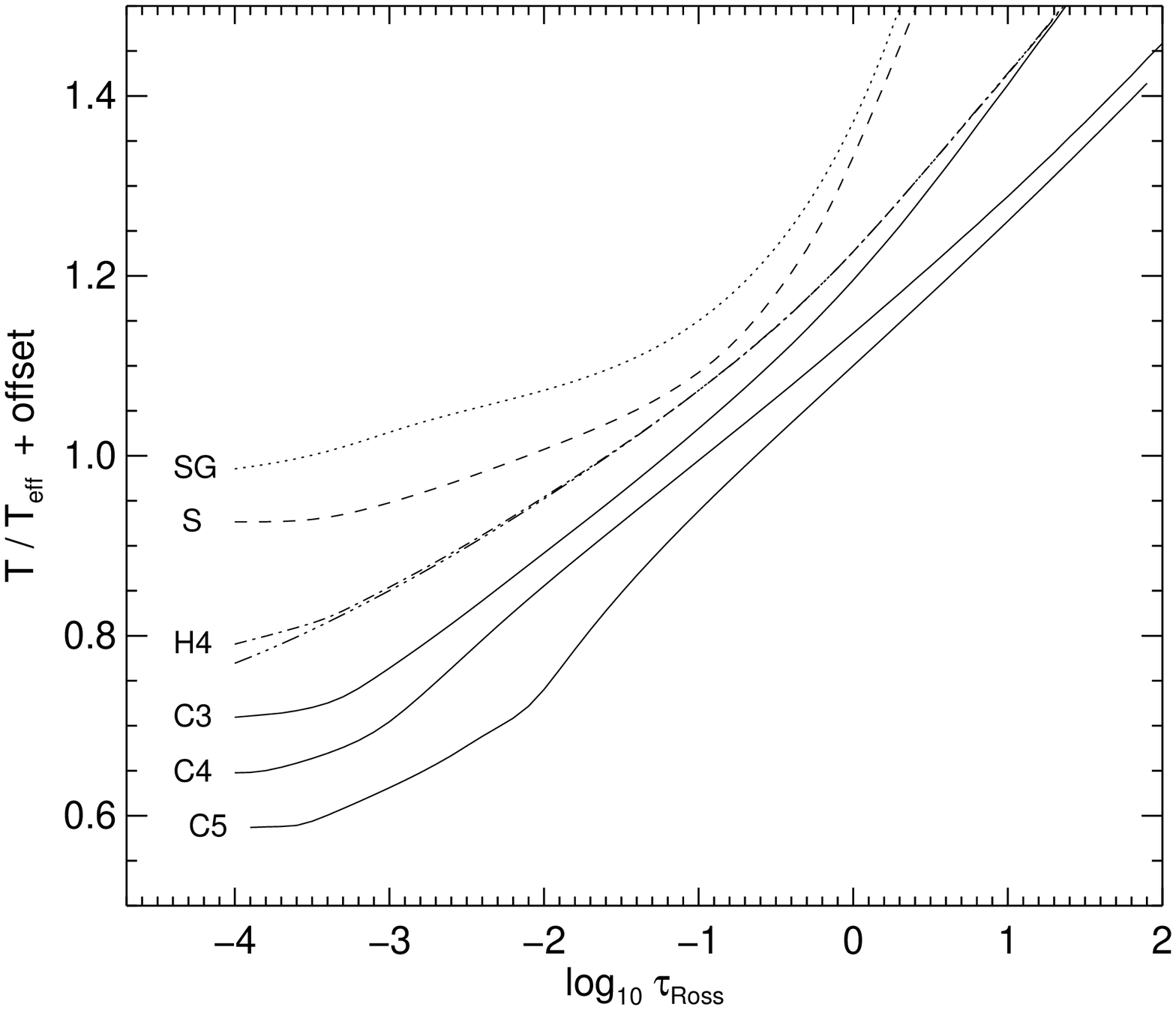}}
\caption[]{Mean temperature (averaged over time and surfaces of constant
  optical depth) relative to effective temperature for the RHD
  models as function of Rosseland optical depth. For clarity the curves are
  shifted by $\{1,2,3,4,5\}\times 0.05$ for the models \mocm, \mocl, \mohm/\mohx,
  \mosu, and \mosg, respectively.}
\label{f:temps}
\end{figure}

The width of the inter-granular lanes relative to the typical granular size
is smaller in our M-type objects of higher gravity. Inspecting the velocity
field (not shown) in vicinity of the continuum forming layers shows less
pronounced size differences. This indicates that the relatively broader lanes
in the solar case are the result of a stronger smoothing of the temperature
field due to a more intense radiative energy exchange, i.e., the effectively
smaller P{\'e}clet number of the flow around optical depth unity in the
hotter objects.  Figure~\ref{f:temps} shows an overview of the mean T-$\tau$
(the temperature was averaged over time and surfaces of equal optical depth)
relations found in the RHD models. One recognizes that in the M-type objects
the thermal structure is influenced by convection to much lower optical depths
than in the solar-type stars.
 
\changed{The temperature model~\mohx\ is on the scale of the plot identical to
  model~\mohm\ in deeper layers but shows a noticeable deviation at low optical
  depth close to the upper limit over which the stratifications have been
  averaged. While perhaps not surprising considering the different placement
  of its upper boundary, the difference may be in part traced back to the
  different viscosity which damps the velocity field and leads to less
  convective heating in the atmospheric layers which are convectively unstable
  (see Sect.~\ref{s:convection} for a discussion of the interplay of
  convection and radiation in the atmospheric layers).}

\subsection{Horizontal scales}

\changed{Primarily due to the variation of the gravitational acceleration (by a factor
of 100) the convective cells in our models span a substantial range in
geometrical size. However, taking the pressure scale height at optical
depth unity~\Hpsurf\ as reference the variation is largely reduced. The
bars in Fig.~\ref{f:grancomp} are placed to suggest that the horizontal size
of the cells indeed roughly scales with~\Hpsurf.}
This is quantified in Fig.~\ref{f:pi} for the intensity
pattern and Fig.~\ref{f:pv} for the pattern of the vertical velocity. The
maximum power in the spectra of the intensity pattern of the M-type objects
lies between 5 and 8\,\Hpsurf, the maximum power of the velocity pattern
between 3 and 6\,\Hpsurf. In both cases the spectra of the solar-type objects
are slightly but noticeable shifted towards smaller wavenumbers (larger spatial
scales). While the relation between M-type and solar-type objects is the same
in both diagnostic variables, it came as a surprise -- at least to the authors
-- that both variables do not provide the same value for the typical cell
size. The different slopes in the intensity and velocity spectra towards
larger scales might be related to this finding. If one considers a pure random
pattern of a given characteristic scale, a slope of unity (in the chosen
representation of power) is to be expected towards larger scales. In this case
the signal at large scales is the result of a mere random superposition of
residual contributions from smaller scale features. The intensity spectra
follow this random model quite closely while the velocity spectra show
noticeably larger deviations with steeper slopes. We do not have an
explanation at hand.  However, considering the range of stellar parameters
covered by our models, the shape of spectra show a large degree of similarity.
In particular, the typical granular scales (in intensity) turn out to be the
same within a factor of two lying between 5 and 10\,\Hpsurf for all objects.

\begin{samepage}
\begin{figure}[!t]
\resizebox{\hsize}{!}{\includegraphics[draft = \draftflag]%
{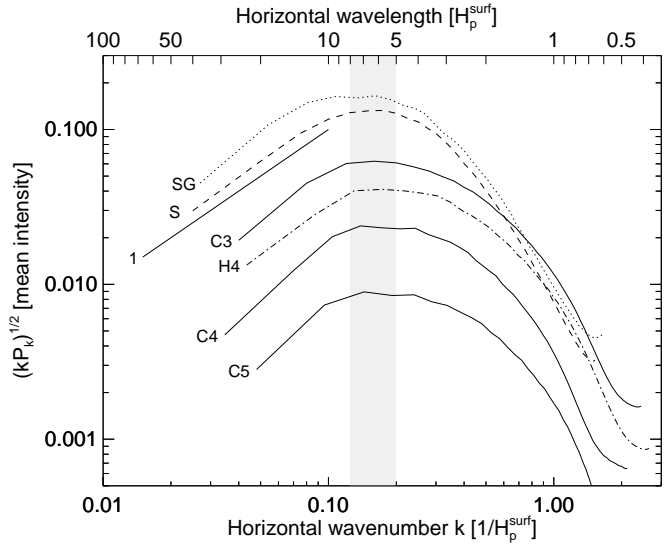}}
\caption[]{Spatial power spectra of the emergent intensity as a
  function of horizontal wavenumber~$k$. The amplitude power is given in
  units of the temporally and horizontally averaged
  intensity. The curves are labeled with the IDs of the models (see
  Tab.~\ref{t:mhdmodels}), the curve labeled ``1'' indicates a
  power-law with slope unity. The grey stripe indicates the range
  of horizontal scales within which the power maxima of the M-type models
  are located.}
\label{f:pi}
\end{figure}

\begin{figure}[!t]
\resizebox{\hsize}{!}{\includegraphics[draft = \draftflag]%
{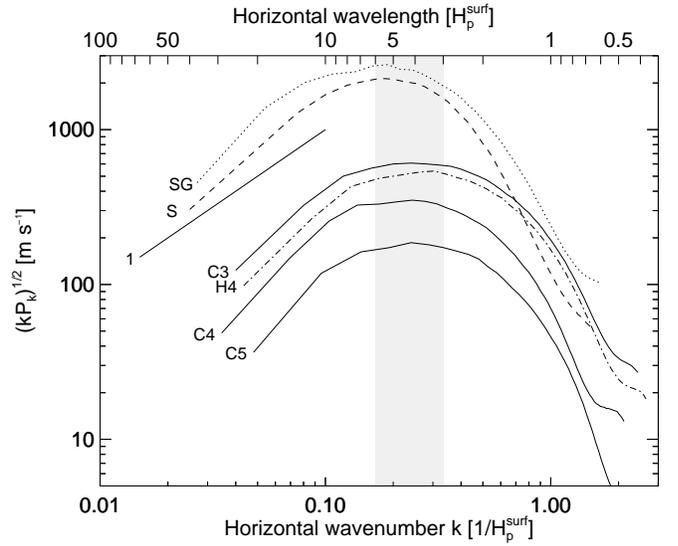}}
\caption[]{Spatial power spectra of the vertical velocity as a
  function of horizontal wavenumber~$k$. The velocities are taken from the
  layer where the convective velocities reach their
  maxima. The curves are labeled with the IDs of the models (see
  Tab.~\ref{t:mhdmodels}), the curve labeled ``1'' indicates a
  power-law with a slope of one. The grey stripe indicates the range
  of horizontal scales within which the power maxima of the M-type models
  are located.}
\label{f:pv}
\end{figure}
\end{samepage}

\subsection{Velocities and turbulent pressure}

Figures~\ref{f:vrms} and~\ref{f:pturb} show the run of the root-mean-square (RMS)
vertical as well as horizontal velocity component, and of the turbulent
pressure, respectively. The averages were taken over time and fixed
geometrical height. As is evident from the figures, the vertical velocity and
turbulent pressure follow each other rather closely. In the higher atmospheric
layers, the velocity field -- especially in the models with more vigorous
convection -- is dominated by horizontal motions. \changed{Test model~\mohx\ closely
follows model~\mohm\ in the deeper layers but shows reduced amplitudes in the higher
layers. As indicated previously we interpret this as primarily a consequence of the
increased numerical viscosity in model~\mohx, and not so much due to the larger
extend of the model.}
Generally, the velocities are smaller in the M-type objects then in the
solar-like objects. Due to their substantially lower \Teff\ in comparison to
the solar-type models, the requirement to transport the nominal energy flux is
already met by convection at lower velocities in the M-type models, leading to
the overall lower ``hydrodynamic activity'' in these objects.

\begin{samepage}
\begin{figure}
\resizebox{\hsize}{!}{\includegraphics[draft = \draftflag]%
{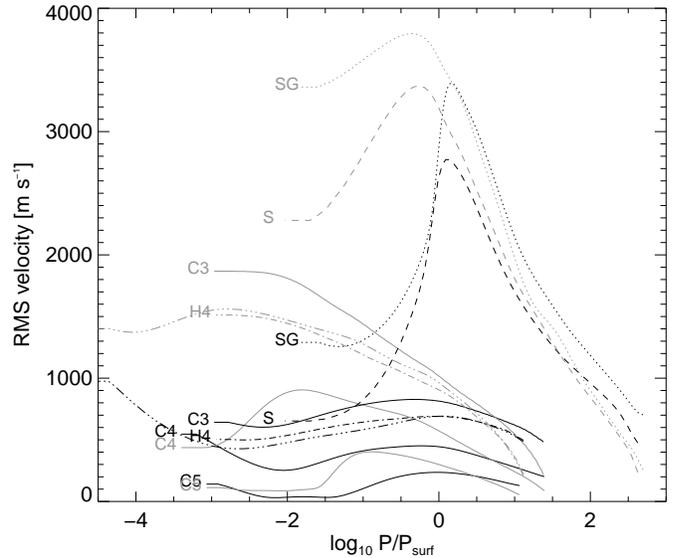}}
\caption[]{Root-mean-square velocities for the RHD models
  as a function of gas pressure relative to the individual surface
  pressures~\Psurf\ (see Tab.~\ref{t:mhdmodels}). Black lines depict
  the vertical, grey lines the horizontal velocity component.}
\label{f:vrms}
\end{figure}
\begin{figure}
\resizebox{\hsize}{!}{\includegraphics[draft = \draftflag]%
{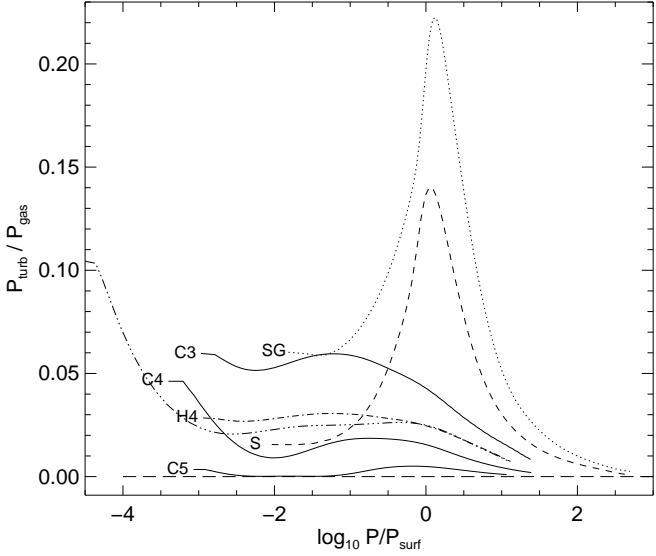}}
\caption[]{Turbulent pressure $\left\langle\rho v_z v_z \right\rangle$ as
  a function of gas pressure relative to the individual surface
  pressures. The turbulent pressure quite closely follows the vertical
  velocity as depicted in Fig.~\ref{f:vrms}.}
\label{f:pturb}
\end{figure}
\end{samepage}

\subsection{Horizontal fluctuations}

Figures~\ref{f:dtrms} and~\ref{f:dprms} show the relative \changed{spatial and
temporal} RMS
fluctuations of temperature and pressure at given geometrical height. In the
M-type models the temperature fluctuations stay at a very modest level nowhere
exceeding 4\,\%\changed{ -- even including model~\mohx}. In the optically thin
layers, model~\mocm\ shows systematically larger temperature fluctuations than
model~\mohm\ of higher \Teff\ but of the same surface gravity. This is an
imprint of the reduced capacity of the radiation field in model~\mocm\ to
smooth horizontal temperature differences.  The pressure fluctuations reach
larger values than the temperature fluctuations, and show a systematic
increase with height.  Note that in our dwarf model~\moch, the pressure
fluctuations only reach a very modest level of about 4\,\%. From this we
expect that in cooler, dust-forming main-sequence objects, thermodynamic
fluctuations are even smaller so that dust formation conditions vary little in
a given layer.
\changed{Model~\mohx\ shows a rapid increase of fluctuations with height
  which is typically found when the flow field is dominated by wave motions.
  The correspondence to model~\mohm\ in the overlapping region is quite good
  indicating that fluctuations in temperature and pressure are little affected
  by the location of the upper boundary.}

\begin{samepage}
\begin{figure}
\resizebox{\hsize}{!}{\includegraphics[draft = \draftflag]%
{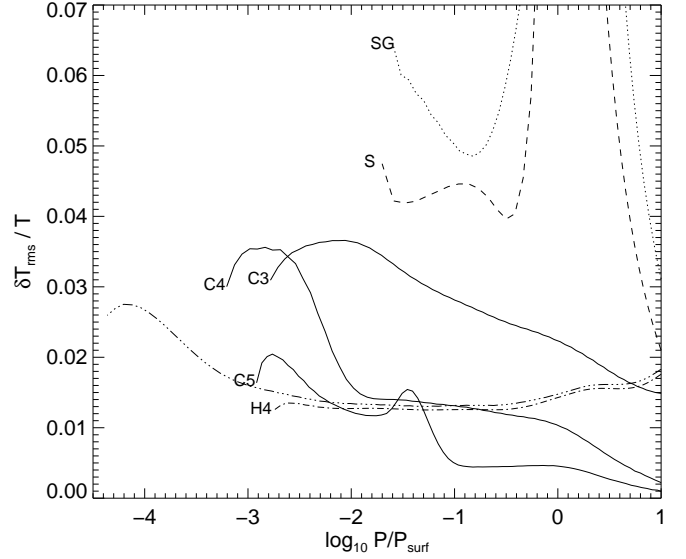}}
\caption[]{Relative RMS temperature
  fluctuations in horizontal planes as a function of gas pressure relative to
  the individual surface pressures~\Psurf\ (see Tab.~\ref{t:mhdmodels}). The
  clipped maxima in the curves for models \mosu\ and \mosg\ reach 0.23 and
  0.30, respectively.}
\label{f:dtrms}
\end{figure}
\begin{figure}
\resizebox{\hsize}{!}{\includegraphics[draft = \draftflag]%
{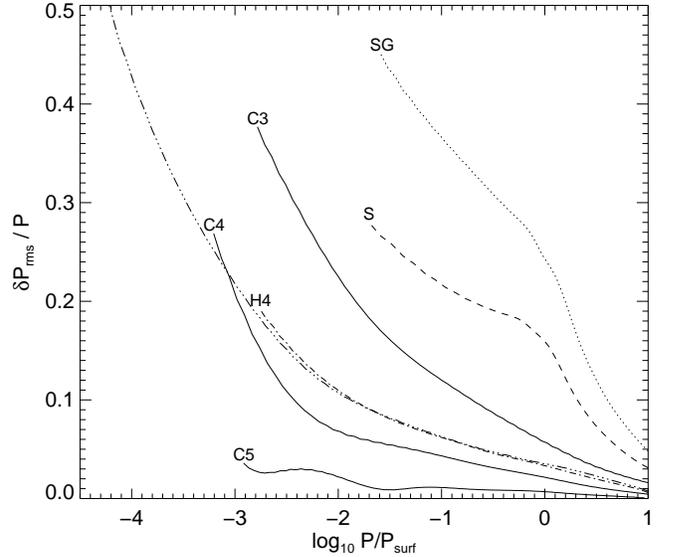}}
\caption[]{Relative RMS fluctuations in horizontal planes 
  as a function of gas pressure relative to the individual surface
  pressures.}
\label{f:dprms}
\end{figure}
\end{samepage}

\subsection{Spatial correlation of vertical velocity and entropy}

In this section, we want to give an overview of some two-point correlations
found in our models. The width of two-point correlations in vertical direction
has been considered as measure of the mean-free-path of mass elements entering
MLT -- the mixing-length~\lmix, and as such has been the target of many
investigations. The (linear) correlation coefficient of a quantity~$x$ between
two layers located  at heights~$z_1$ and $z_2$ is given by
\beq
\cov{x_1}{x_2} = \frac{\xtmean{x_1 x_2}-\xtmean{x_1}\xtmean{x_2}}{\sig{x_1}\sig{x_2}}
\label{e:twopointcorr}
\eeq
where \sig{x_i} is the standard deviation of $x$ at height~$z_i$, and the
angular brackets denote the average over time and horizontal position. In a
seminal paper, \citet{Chan+Sofia87} found in hydrodynamical models of stratified
{\em efficient\/} convection that the full-width-at-half-maximum (FWHM) of the
correlation function of the vertical velocity, as well as temperature, scaled
with the pressure scale height and not density scale height.  In a follow-up
study, \citet{Chan+Sofia89} showed that the result is robust against variations
of the ratio of the specific heats $\gamma$ of the gas. \citet{Singh+Chan93}
found that the width of correlation changes moderately with Prandtl number.
\citet{Kim+al95} studied the case of {\em inefficient\/} convection in the
Sun. Their model included radiative transfer effects (in diffusion
approximation) as well as effects of the ionization of hydrogen and helium.
Under the conditions studied by Kim and collaborators, the width of the
correlation function of the velocity scaled with pressure {\em and\/} density
scale height, while the width of the temperature correlation with neither of
both.  \citeauthor{Kim+al95} had to restrict their investigation to optically
thick regions. \citet{Robinson+al03,Robinson+al04} included also optically
thin layers in their hydrodynamical models of the Sun and a subgiant star (at
\Teff=4990\pun{K}, \logg=3.37). The radiative transfer was treated in grey
Eddington approximation.  Robinson and collaborators found a complex
height-dependence of the width of the correlation of the vertical velocity and
entropy fluctuations. Especially the last result shows that we cannot expect
to find a universal behavior of the correlations in stratified convection
under all possible circumstances -- in particular, in the case of inefficient
convection which is most important for stellar structure models. However,
there might be still hope to find general trends which may serve as buildings
blocks for an improved treatment of convection beyond MLT.

\changed{Figures~\ref{f:vzcorr1} and~\ref{f:vzcorr2} provide an overview of the
velocity and entropy correlations found in our models. Our models contribute
to the ongoing discussion in a twofold way: they cover a large range of
stellar parameters and include an elaborate treatment of the radiative
transfer in the optically thin layers.  The overall behavior of the
correlations is complex, but some general features can be identified: i)
considering that the models cover about a factor of two in \Teff\ and one
hundred in \logg, a certain uniformity in the width of the distributions in
the convection dominated layers is apparent; ii) in the sub-photospheric
layers the entropy correlation is more peaked than the velocity correlation;
iii) with the exception of model~\moch\ the velocity correlation become
broader in radiation dominated layers; iv) the widths of the correlations tend
to shrink with increasing depth, possibly towards an asymptotic limit
(different for velocity and entropy) -- in line with the findings of
\citet{Chan+Sofia89}; v) the width of the entropy correlation of the solar
model~\mosu\ and subgiant~\mosg\ passes through a pronounced minimum around
optical depth unity; this is accompanied by anti-correlations signifying the
thermal behavior of overshooting motions; vi) the degree of uniformity of the
correlations is not improved if plotted on the density scale.

The limiting width of the correlation functions is fairly well defined and
given in the panels of Figs.~\ref{f:vzcorr1} and~\ref{f:vzcorr2}. The width of
the velocity correlation lies in the same range as values of the mixing-length
parameter associated with certain model features which we are going to discuss
later.  However, one must keep in mind that the width of the correlations
varies substantially -- sometimes even dramatically -- in the models, and that
the widths are different for different quantities. In our opinion, a direct
association of a width of a correlation with a mixing-length parameter is an
over-simplification of the actual situation.


\begin{figure}
\resizebox{\hsize}{!}{\includegraphics[draft = \draftflag]%
{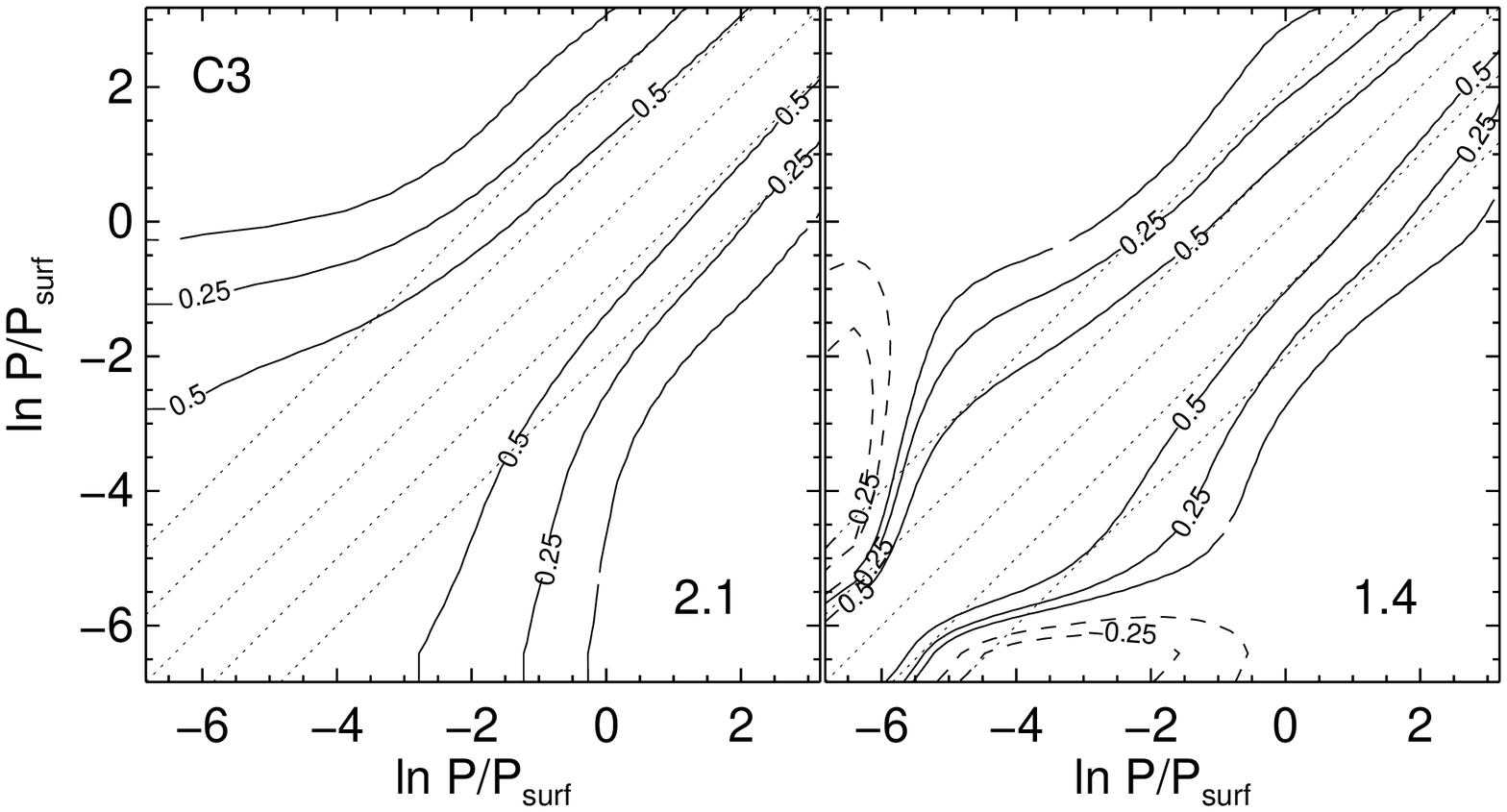}}
\resizebox{\hsize}{!}{\includegraphics[draft = \draftflag]%
{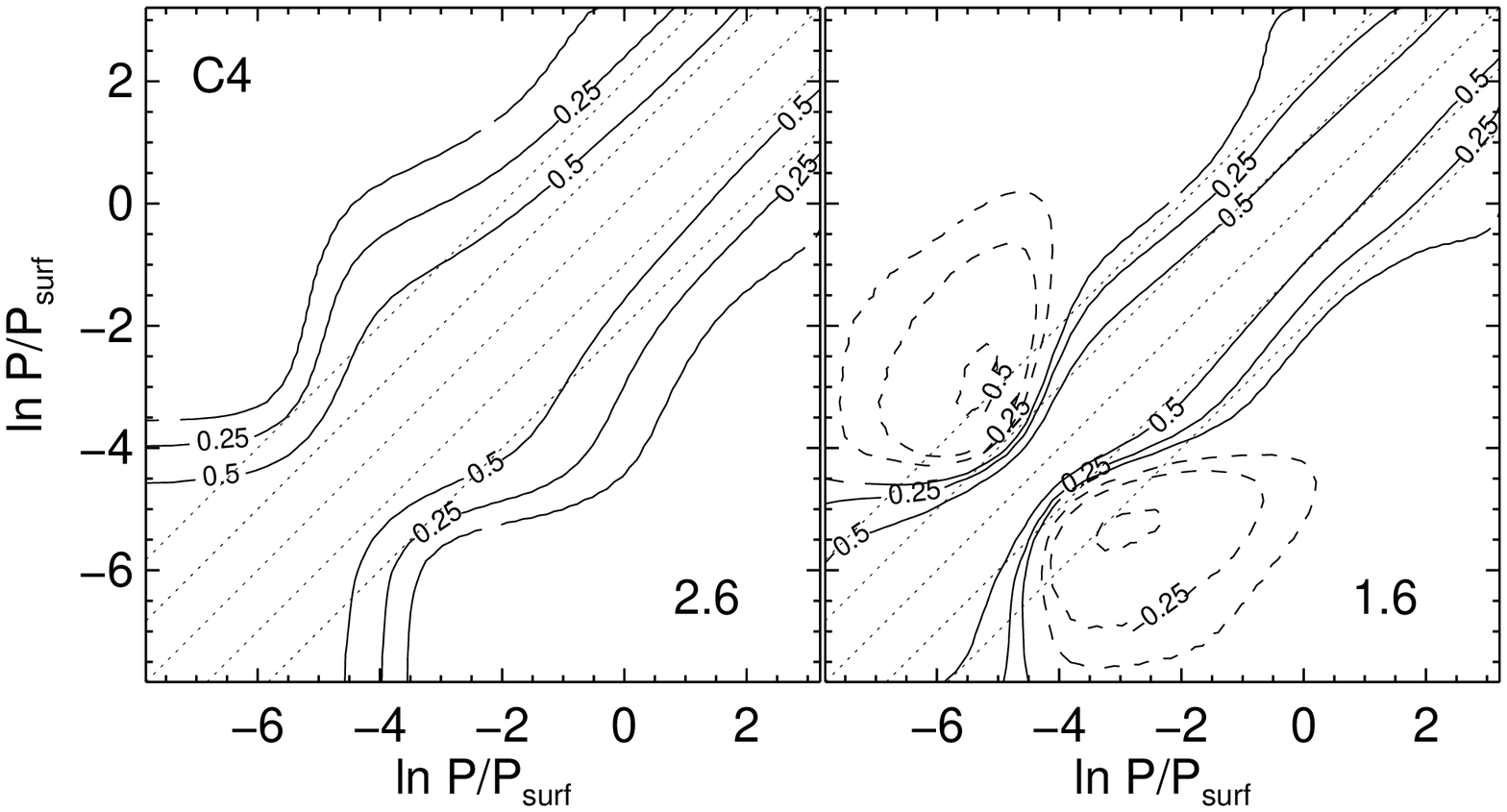}}
\resizebox{\hsize}{!}{\includegraphics[draft = \draftflag]%
{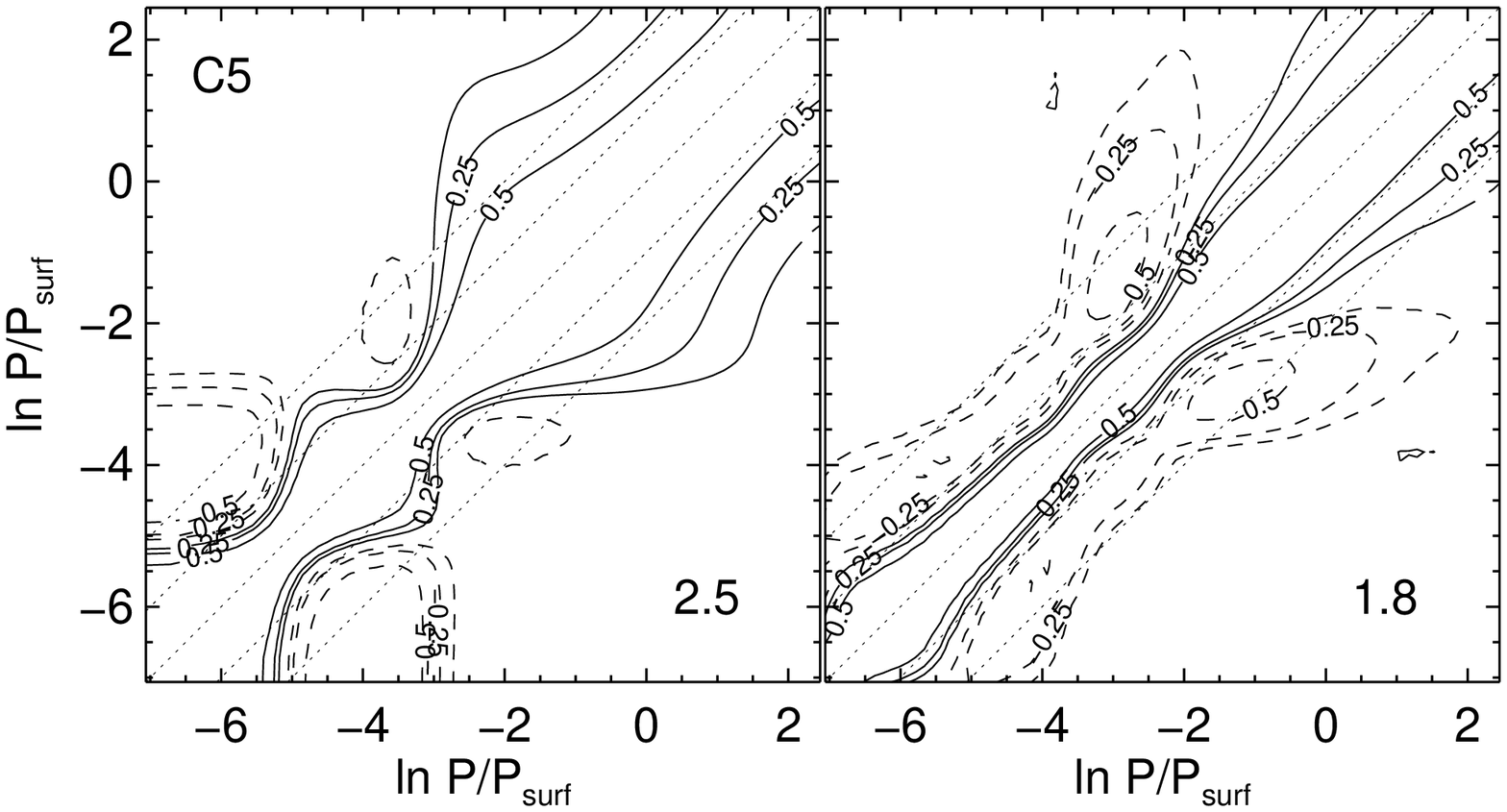}}
\caption[]{Contour plots of the two-point spatial correlation functions of the
  vertical velocity (left panels), and entropy (right panels) for the models
  \mocl, \mocm, and \moch. The pressure relative to the surface pressure of
  the two involved height levels (cf. Eq.~\eref{e:twopointcorr}) are given on
  abscissa and ordinate. The height coordinates are interchangeable due to the
  symmetry of the correlation function. The number in the lower right corner
  of each panel gives the FWHM of the correlation (in units of the
  \textit{local} pressure scale height \Hp) in vicinity of the
  lower boundary of the computational domain. Contour lines are given for
  values of $\{-0.5, -0.25, -0.125, 0.125, 0.25, 0.5\}$. Lines with positive
  contour values are depicted by solid lines, negative ones by dashed lines.
  Dotted lines are for orientation and have a distance of $\Delta\ln(P/\Psurf)
  = 1$. The central dotted line coincides with the maximum of the correlation
  function which is normalized to one.}
\label{f:vzcorr1}
\end{figure}
}

\begin{figure}
\resizebox{\hsize}{!}{\includegraphics[draft = \draftflag]%
{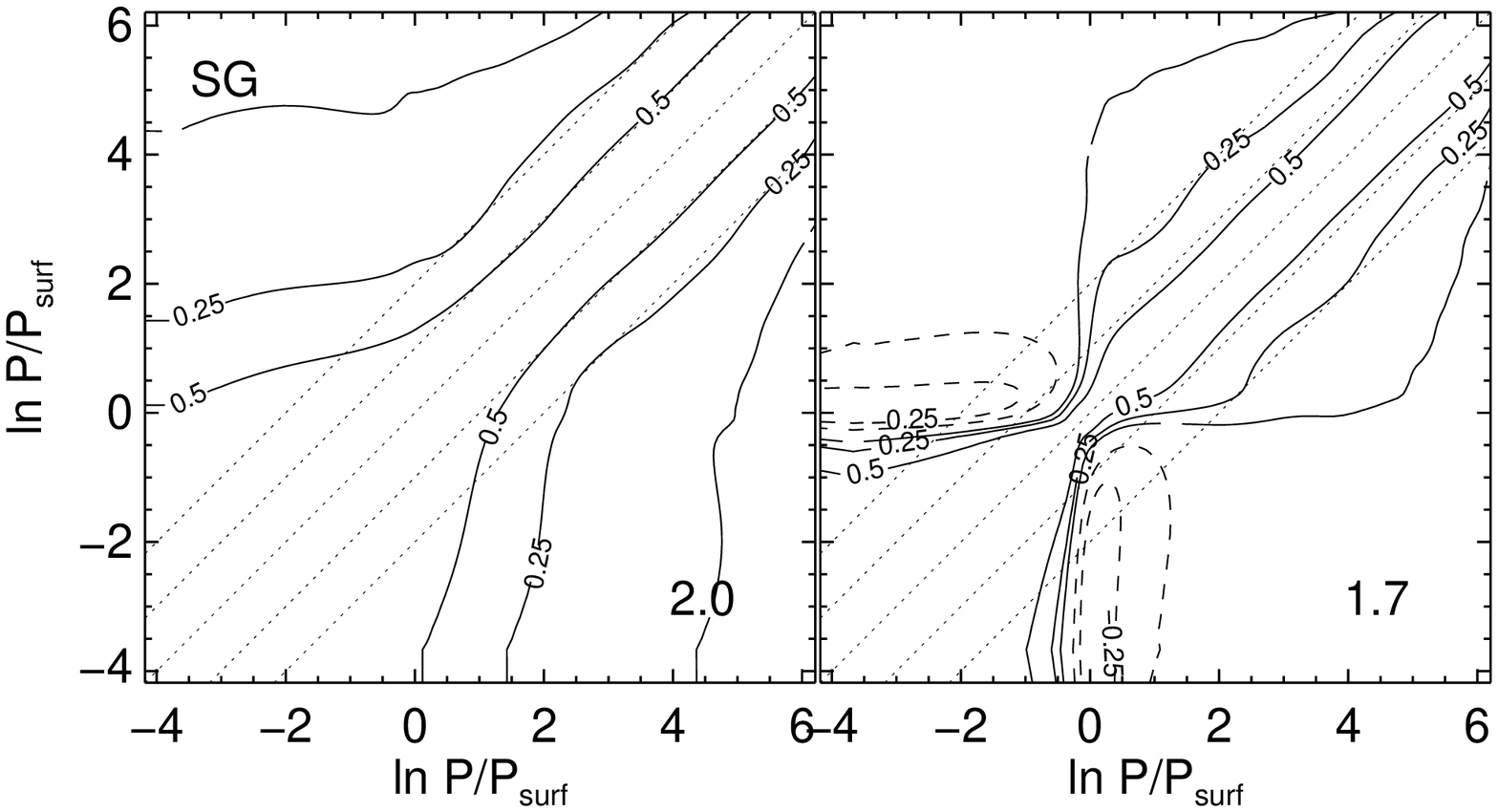}}
\resizebox{\hsize}{!}{\includegraphics[draft = \draftflag]%
{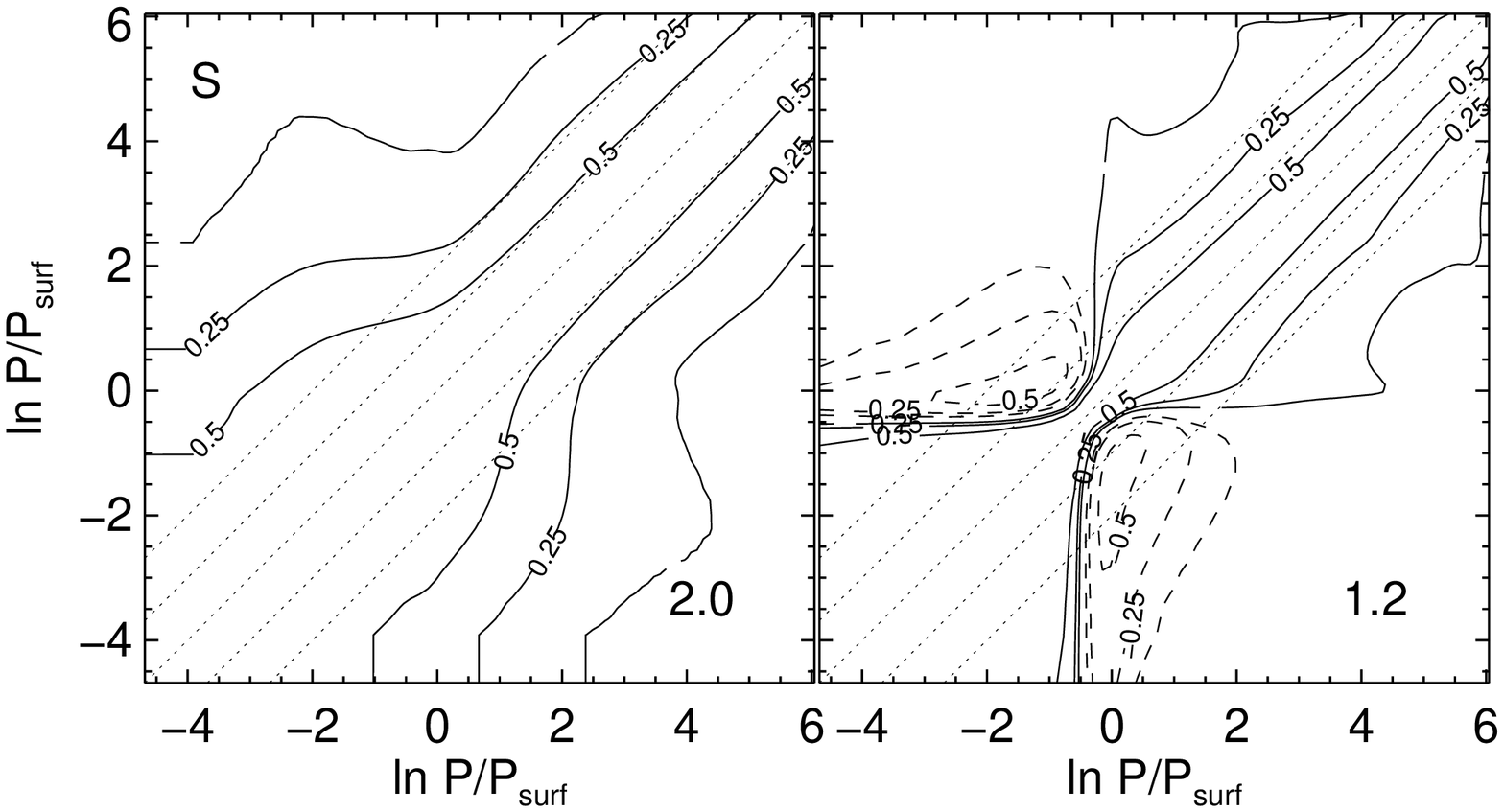}}
\resizebox{\hsize}{!}{\includegraphics[draft = \draftflag]%
{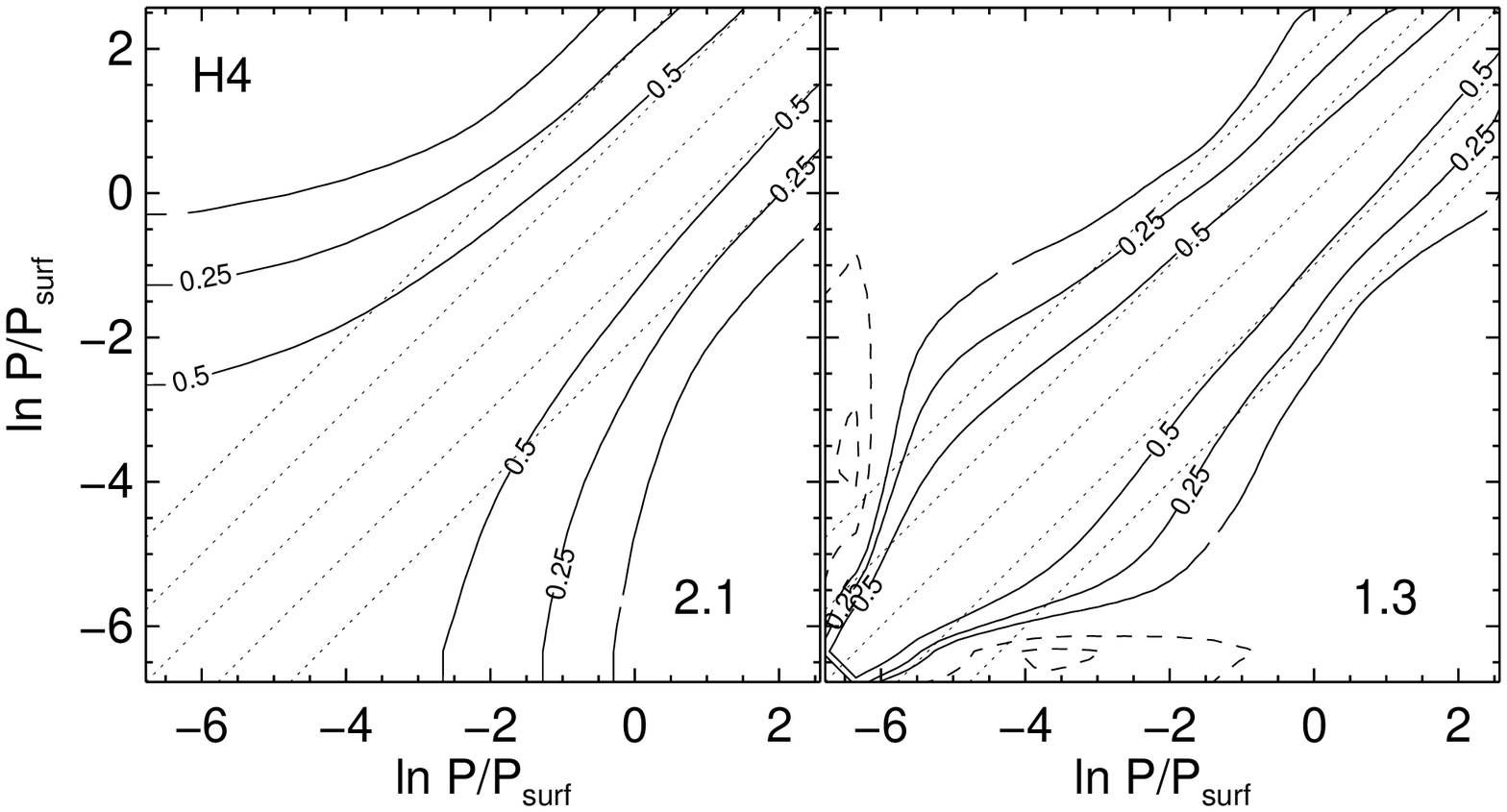}}
\caption[]{Same as Fig.~\ref{f:vzcorr1} for models \mosg, \mosu, and \mohm.}
\label{f:vzcorr2}
\end{figure}

\section{Convective energy transport}
\label{s:convection}

Convection is an important energy transport mechanism in M-type stars. In
standard model atmospheres it is treated in the framework of MLT. In this
section, we want to address the question whether the simplistic MLT is actually
capable to provide a sufficiently accurate description of the convective
energy transport under conditions encountered in M-type atmospheres.
Figure~\ref{f:entropy} shows a comparison of the entropy structure of the
RHD model atmospheres and standard 1D hydrostatic models in
radiative-convective equilibrium assuming different mixing-length parameters.
Figure~\ref{f:temps2} depicts corresponding temperature profiles which allow to
approximately translate entropy differences among the profiles in
Fig.~\ref{f:entropy} to temperature differences.

To calculate the entropy profiles of the RHD models, they have been
averaged temporally and horizontally on surfaces of constant optical depth.
This procedure ensures a particularly good preservation of the energy
transport properties of the RHD models \citep{Steffen+al95}.
Moreover, it reduces the ``smearing'' of vertical gradients by plane-parallel
oscillations which occurs when averaging over fixed horizontal planes.
Besides the mixing-length parameter itself, MLT contains a number of further
``hidden'' parameters intrinsic to the specific formulation of MLT which was
chosen. We emphasize that a well-defined calibration of the mixing-length
parameter must always be given with reference to the specific formulation in
operation.  Here, we are using the formulation given by \citet{Mihalas78}. See
\citet{Ludwig+al99} for details of the implementation.

As already remarked earlier, Fig.~\ref{f:entropy} illustrates that the
sensitivity of the structure of the standard models to the mixing-length
parameter increases with decreasing gravity as well as increasing \Teff.
Model~\mocl\ shows a sensitivity of the entropy in the deep, adiabatic layers
which is comparable to the sensitivity of solar MLT
models. Figure~\ref{f:entropy} also illustrates that the convectively
unstable layers (with entropy gradient $\frac{ds}{d\tau}>0$) extend to
small -- spectroscopically important -- optical depths. 
M-type atmospheres offer the opportunity to
study convection under optically thin conditions. 

\begin{samepage}

\begin{figure}[!t]
\resizebox{\hsize}{!}{\includegraphics[draft = \draftflag]%
{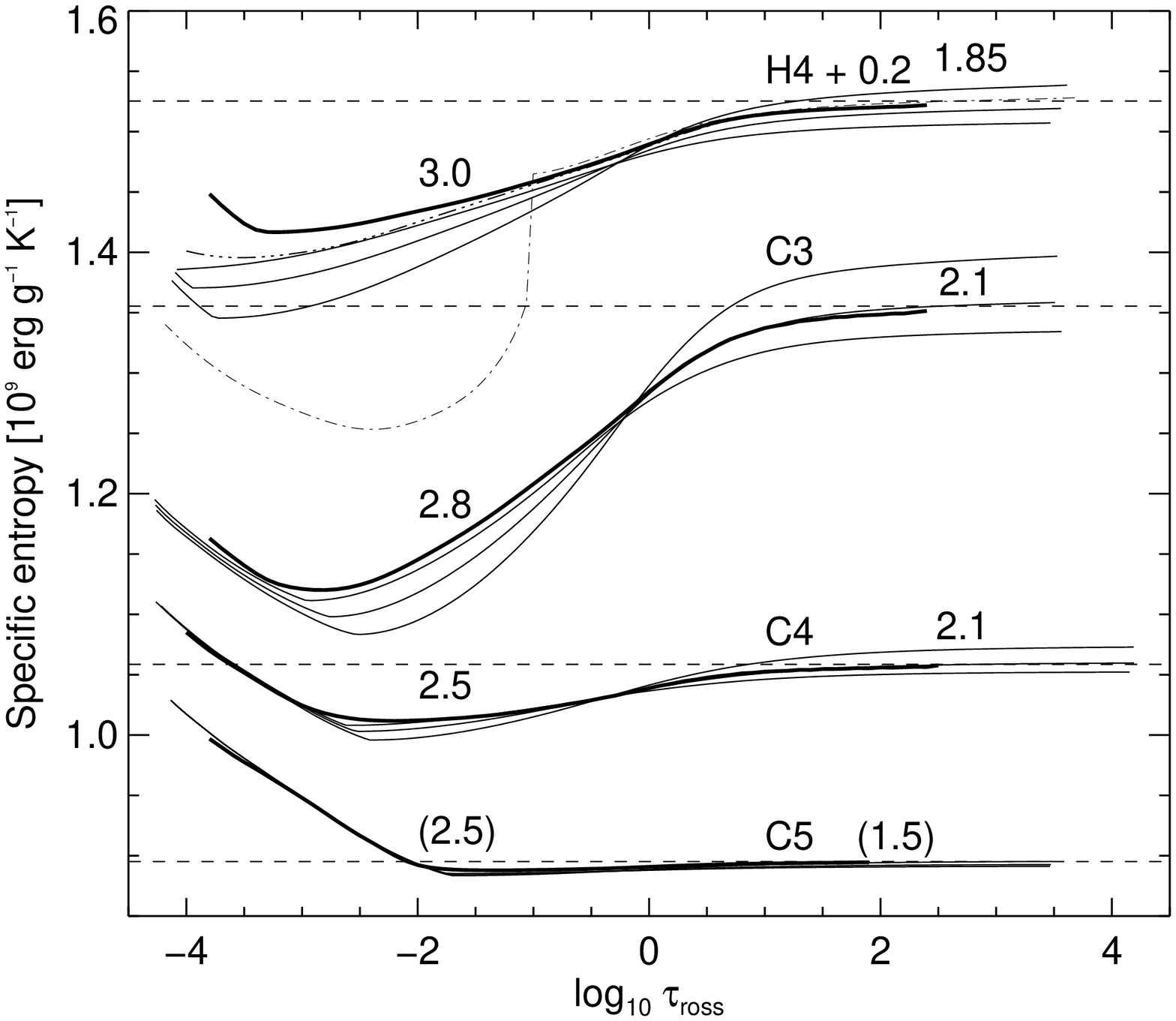}}
\caption[]{Entropy as a function of Rosseland optical depth of the RHD models (thick solid
  lines) in comparison to standard mixing-length models (thin solid lines).
  For each RHD model three MLT models are plotted with \mlp=1.5,
  2.0, and 2.5. The entropy of the MLT models behaves monotonically with \mlp,
  the \mlp=1.5 model having the lowest entropy in the optically thin, and the
  highest in the optically thick layers. Model H4 has been offset by $+0.2$
  entropy units for clarity, and one special MLT model with \mlp=2.0 has been
  added (dashed-dotted line; see text). The dashed lines depict the value of
  entropy present in the adiabatically stratified regions of the convective
  envelope. Numbers indicate mixing-length parameters necessary to match the
  RHD structure by MLT models.}
\label{f:entropy}
\end{figure}

\begin{figure}[!t]
\resizebox{\hsize}{!}{\includegraphics[draft = \draftflag]%
{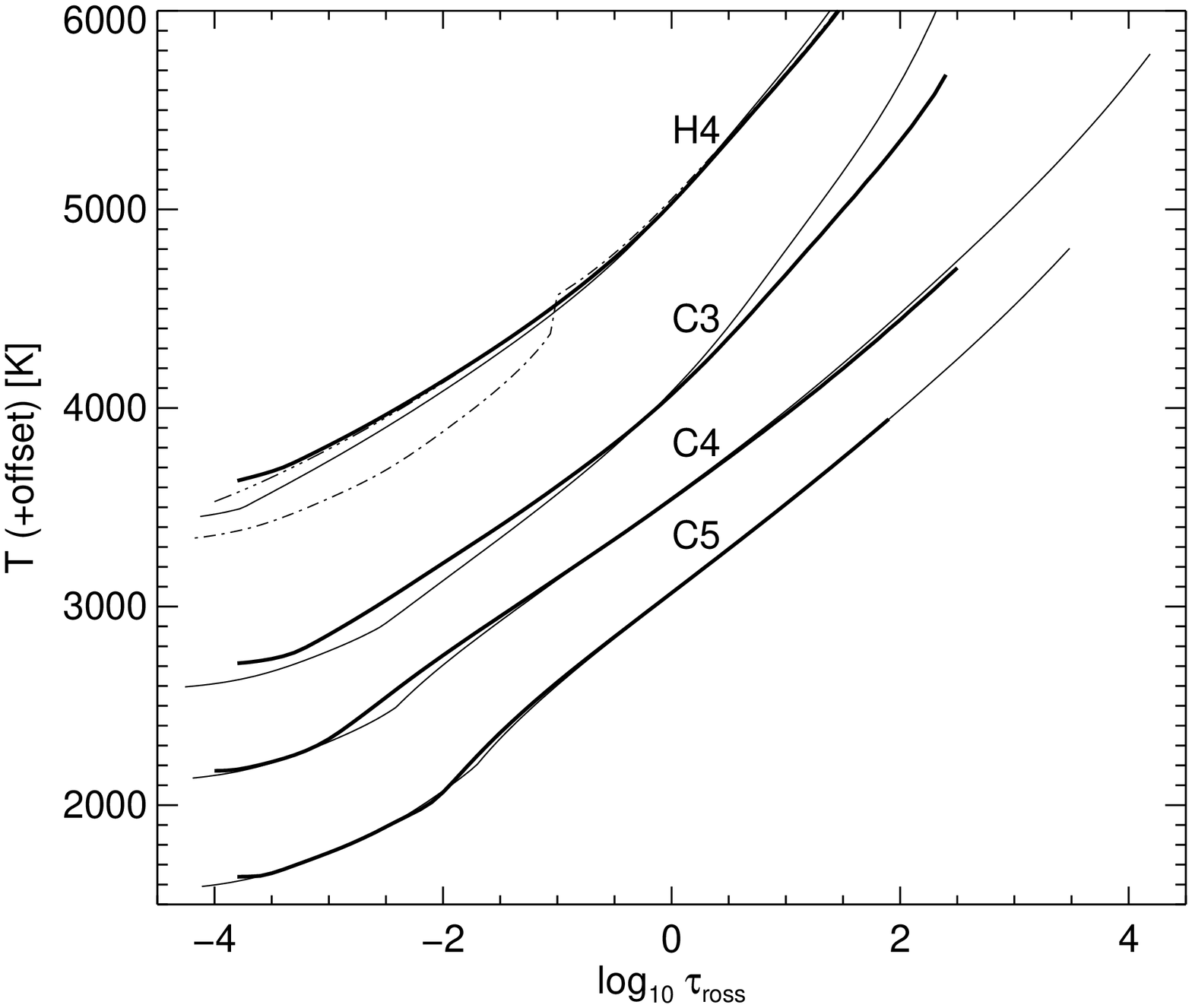}}
\caption[]{Temperature as a function of Rosseland optical depth of the 
  RHD models (thick solid lines) in comparison to standard
  mixing-length models (thin solid lines). The MLT models are calculated
  assuming \mlp=1.5. For model H4 a special MLT model with \mlp=2.0 has been
  added (dashed-dotted line; see text). For clarity, models C4, C3, and H4
  have been offset by 500, 1000, and 1500\pun{K}, respectively. }
\label{f:temps2}
\end{figure}

\end{samepage}

\begin{figure}
\resizebox{\hsize}{!}{\includegraphics[draft = \draftflag]%
{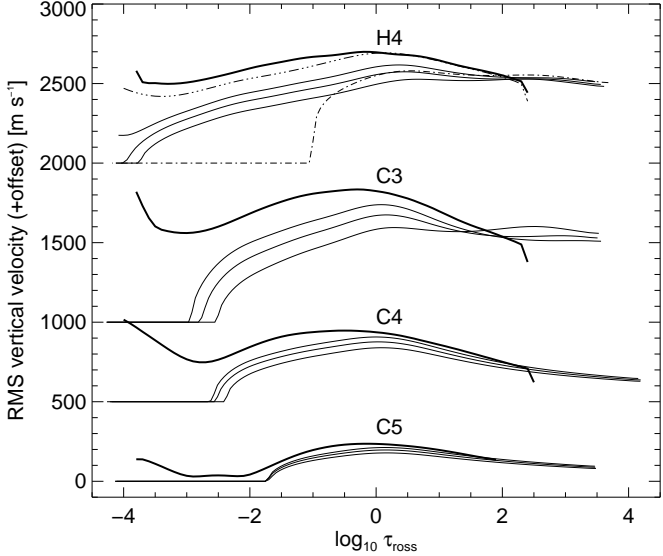}}
\caption[]{RMS vertical velocity component of the
  RHD models (thick solid lines) in comparison to convective
  velocities from MLT models with \mlp=1.5, 2.0, and 2.5 (thin solid lines) as
  a function of optical depth. The velocities from MLT increase in the
  optically thin layers monotonically with increasing \mlp. For model H4 a
  special MLT model with \mlp=2.0 has been added (dash-dotted line; see text).
  Models C4, C3, and H4 have been shifted by 500, 1000, and
  2000\pun{m\,s$^{-1}$}, respectively. The downturn of the velocity in models
  C4, C3, and H4 at largest optical depth is an artifact of the averaging
  procedure and should be ignored.}
\label{f:velmlt}
\end{figure}

The general role of convection can be described as follows: in the
convectively unstable layers -- here comprising also parts of the optically
thin layers -- the thermal structure is the result of a {\em competition
  between adiabatic heating and radiative cooling\/} because a temperature
structure in adiabatic equilibrium would be hotter than in radiative
equilibrium. The mixing-length model (with $\mlp=2.0$) depicted by the
dashed-dotted line for case~\mohm\ is intended to illustrate this (see
Figs.~\ref{f:entropy}, \ref{f:temps2}, and~\ref{f:velmlt}): in this model we
artificially switched off the convective motions in the layers with
$\log\tauross\leq -1$. This suppresses the convective heating and forces the
temperature to adjust to radiative equilibrium conditions. As evident from the
figures, this leads to a substantial drop of entropy and temperature.  In the
convectively stable layers (with entropy gradient $\frac{ds}{d\tau}<0$) the
situation is reversed, and the temperature is controlled by a {\em balance
  between adiabatic cooling and radiative heating\/} -- as far as the
RHD models are concerned. The situation is different for the MLT
models where by construction no convective (overshooting) motions take place
in the formally stable layers, and the temperature is determined by the
condition of radiative equilibrium alone.

For decreasing \logg\ at given \Teff, the models tend to stay closer to
radiative equilibrium conditions in the optically thin layers. Two factors
reduce the efficiency of the convective energy transport: first,
Fig.~\ref{f:kappa} shows that the opacity does not vary radically among the
M-type models. Hence, lower gravity models exhibit lower densities at given
optical depth, which reduces the thermal energy that can be transported per
unit volume, rendering the convective transport of heat more difficult.
Second, the pressure scale height increases with decreasing gravity while
typical convective velocities increase only modestly. This increases the time
scale over which vertically traveling mass elements change their temperature
due to adiabatic expansion or compression. The radiative time scale in the
optically thin regions, on the other hand, is independent of the spatial
scales and mass density which again leads to a shift of the thermal balance
towards radiative equilibrium conditions. 

As evident from Fig.~\ref{f:vad}, the pressure-temperature dependence of the
adiabatic gradient counter-acts the trend towards radiative equilibrium
conditions in our M-type models of lower gravity. In models~\mocl\ and~\mohm, 
the adiabatic gradient is close to its minimum favoring convection due to the
formation of H$_2$ molecules. Nevertheless, the appreciable sensitivity of the
MLT models to the mixing-length parameter at lower gravities indicates that
convection and radiation operate with comparable efficiency.

\begin{samepage}
\begin{figure}
\resizebox{\hsize}{!}{\includegraphics[draft = \draftflag]%
{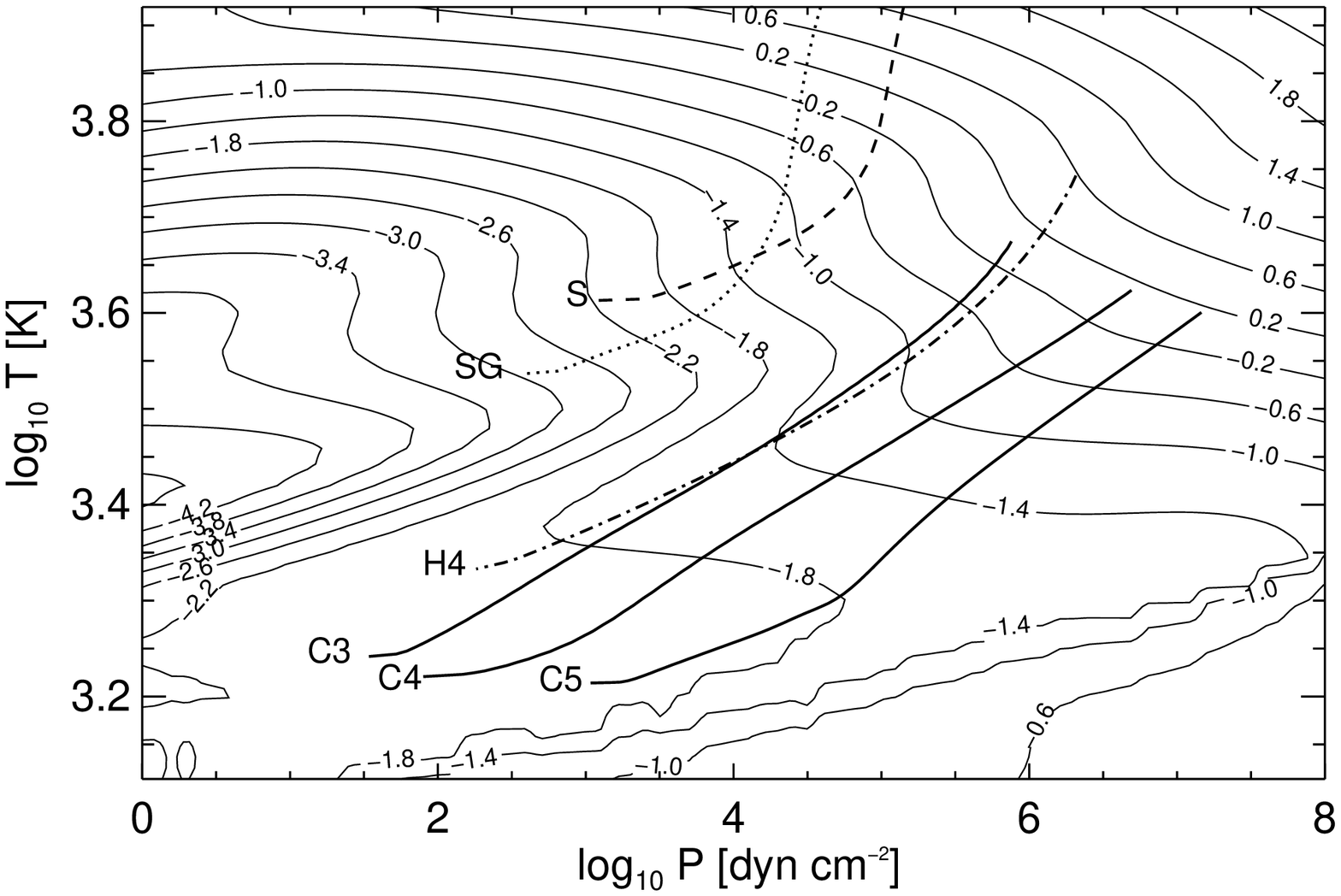}}
\caption[]{Average pressure-temperature profiles of the RHD models
  (thick lines) overlayed on contours (thin lines) of the ($\log_{10}$)
    Rosseland mean opacity.}
\label{f:kappa}
\end{figure}

\begin{figure}
\resizebox{\hsize}{!}{\includegraphics[draft = \draftflag]%
{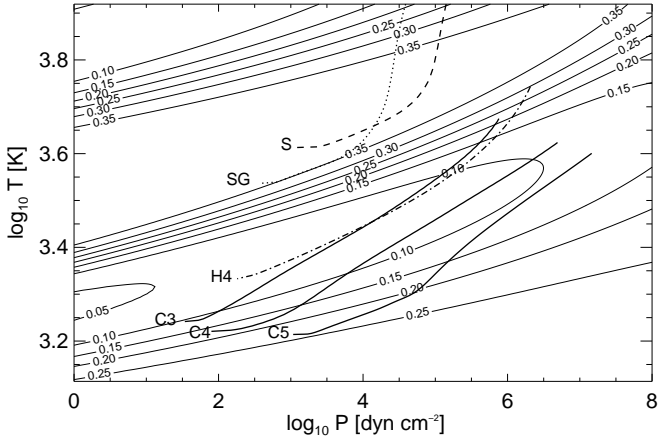}}
\caption[]{Average pressure-temperature profiles of the RHD models
  (thick lines) overlayed on contours (thin lines) of the adiabatic gradient.}
\label{f:vad}
\end{figure}
\end{samepage}

In Figs.~\ref{f:entropy} and~\ref{f:temps2} we compare groups of models of the
same \Teff\ and \logg, and consequently the models in each group are
constrained to similar temperatures and entropies in vicinity of optical depth
unity. Since the entropy- as well as temperature-gradient at this location
depend on the mixing-length parameter, this leads to the crossing of the
various profiles around $\tauross=1$. Deviations from this behavior -- in
particular shown by the RHD models -- come about by residual
differential changes of the opacities and thermodynamic properties of the
matter among the models. In the case of the RHD models, horizontal
fluctuations together with non-linearities in the material functions add to
the deviations.

Figure~\ref{f:entropy} shows that irrespective of the choice of the
mixing-length parameter~\mlp, no MLT model is capable to match the whole
average thermal profile of a hydrodynamical model -- at least within the
framework of the MLT formulation adopted here. We nevertheless give estimates
of the mixing-length parameter matching certain features of the hydrodynamical
structure. Multi-dimensional hydrodynamical convection models provide the
value of the entropy asymptotically reached in the deep, adiabatically
stratified layers of the convective envelope, since it is identical to the
entropy of upflowing material in the deepest regions of the hydrodynamical
model (for a discussion of the scenario underlying this notion see
\citet{Steffen93}, and \citet{Ludwig+al99}). We find a mixing-length parameter
of 1.5, 2.1, 2.1, and 1.85 for models \moch, \mocm, \mocl, and \mohm,
respectively, to match the asymptotic entropy. This \mlp\ is most relevant for
stellar structure models since the asymptotic entropy has a large influence on
the radius of a convective stellar object. The value for model~\moch\ (1.5) is
highly uncertain since the sensitivity to \mlp\ is very small -- however, its
precise value is also not particularly important since the asymptotic adiabat
hardly depends on \mlp. Attributing little weight to model~\moch, we find that
the typical \mlp\ suitable for evolutionary models lies around $\approx 2$ for
mid to late M-type atmospheres. This value is not very different from the
value found for the Sun ($\approx 1.8$), but we do not consider this as an
indication of an universal value for \mlp\ as was already discussed in LAH.

Now we turn to the optically thin parts of the atmosphere relevant for
spectroscopy. We find that mixing-length parameters for matching the
temperature in the range $-2\leq\log\tauross\leq -1$ of 2.5, 2.5, 2.8, and
3.0, for models \moch, \mocm, \mocl, and \mohm, respectively. This range in
optical depth was chosen as representative of the deeper photosphere. However,
the match in this part does not imply a match over the whole optically thin
region. Again, the value for model~\moch\ is uncertain, but not particularly
important. All in all we obtain a range of $\mlp=2.5\ldots 3.0$ when matching
the thermal profile of the RHD models. The temperature structure of the
hydrodynamical models in the uppermost, convectively stable part of the
atmosphere does not deviate much from radiative equilibrium profiles judging
by extrapolating from the available MLT models. This indicates that cooling by
convective overshoot or heating by waves is not very efficient in the layers
immediately adjacent to the surface convective zone.

Model~\mohm\ is an exception from the general trend since it becomes
convectively stable much earlier than one might expect from extrapolating the
MLT models.  The exceptional behavior is due to the fact that in this model
the adiabatic gradient changes very little along the profile in the upper
photosphere (see Fig.~\ref{f:vad}). The same holds for the actual temperature
gradient in the convectively unstable part, making the exact location of the
transition from convective instability to stability very sensitive to the
actual run of the temperature. Here, we have an example where second order
effects can enhance differences to MLT models. Nonetheless, overshooting and
wave heating, again, show little impact on the temperature gradient in the (not
very extended) stable zone.

\changed{A comparison of models~\mohm\ and~\mohx\ in Fig.~\ref{f:entropy}
  reveals that the modest temperature differences in the optically thin layers
  displayed in Fig.~\ref{f:temps2} correspond to a sizable entropy difference
  which corresponds to a decrease of the mixing-length parameter of about 0.5
  relative to model~\mohm. Such a change would bring the atmospheric value of
  the mixing-length parameter of model~\mohm\ closer to the other M-type
  models. As argued before the change in the thermal structure of model~\mohx\
  can only be partially attributed to a systematic influence of the upper
  boundary condition.  Nevertheless, taking the change as an estimate of the
  uncertainty of the derived atmospheric mixing-length parameter, one would
  still find a rather high value of the mixing-length parameter for
  model~\mohm\ of at least 2.5.  The mixing-parameter for matching the
  asymptotic entropy remains the same between models~\mohm\ and~\mohx.}

In LAH we discussed the mixing-length parameter necessary to match the maximum
vertical RMS velocity of the hydrodynamical model~\moch\ and found a value of
3.5. While we do not perform a similar matching here, Fig.~\ref{f:velmlt}
clearly shows that for all RHD models a value substantially larger than 2.5 --
including model~\mohx\ -- is necessary to match the maximum velocity.

Comparing the overall situation that we encounter in M-type objects to stars
of roughly solar effective temperature (and of solar composition) we find that
the transition from convectively to radiatively dominated energy transport
happens more gradually in M-type objects. This is ultimately linked to the
different temperature sensitivity of the dominant opacity (H and H$^-$
bound-free and free-free absorption versus TiO and H$_2$O molecular line plus
H$^-$ bound-free and free-free absorption), and dominant thermodynamic process
(H recombination versus H$_2$ molecular formation) encountered in the two
regimes of effective temperature. The values of \mlp\ we obtained here when
matching the asymptotic entropy are not too different from the values obtained
for solar-type stars. However, in solar-type stars a calibrated MLT model
merely provides the correct entropy jump. The actual run of the entropy in the
optically thick layers is not very well matched: usually, a RHD model predicts
a more rapid switching between adiabatically and radiatively stratified
layers. In M-type objects, a calibrated MLT model matches the actual thermal
profile in the optically thick regions more closely. This property is likely
related to the more gradual transition between the two modes of energy
transport.



\subsection{Spectroscopic effects}

\changed{%
In this section, we want to demonstrate which impact the differences
between the thermal structures of RHD and MLT models have on spectral
properties. \citet{Mohanty+al04a} used molecular bands of
titanium-oxide and lines of neutral atomic alkalis to determine the effective
temperatures and surface gravities of M-type PMS objects by comparing
synthetic and observed spectra. The temperatures and gravities of the objects
studied by Mohanty and collaborators fell into the regime considered here.
The authors emphasized that the strengths of the investigated TiO band heads
serve as excellent and important temperature indicator.  Figures~\ref{f:TiO}
and~\ref{f:NaI} show two prime spectral regions (wavelengths are given as
wavelengths in air) considered in the analysis by Mohanty et al.\,.  For our
comparison, we picked case~\mocl\ where we found significant differences
in the thermal structures between RHD and MLT models. Spectral synthesis
calculations were performed with the PHOENIX code on the prescribed structures
at a spectral resolution of 0.01\pun{\AA}. In Figs.~\ref{f:TiO} 
and~\ref{f:NaI}
the spectral resolution has been degraded to $\sim 30\,000$ similar to the one
in the work of Mohanty et al.\,.

\begin{samepage}
\begin{figure}
\resizebox{\hsize}{!}{\includegraphics[draft = \draftflag]%
{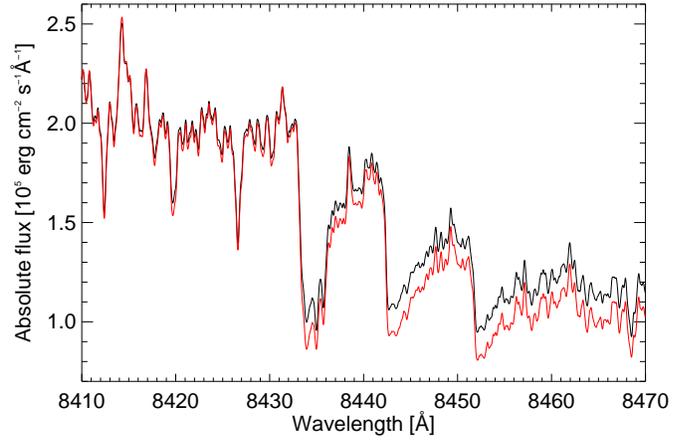}}
\caption[]{Comparison of synthetic spectra showing the triple-headed epsilon
  ($E^3{\Pi}-X^3{\Delta}$) band of TiO (at 8432, 8442, 8452\pun{\AA}), based on
  the hydrodynamical structure~\mocl\ (black solid line) and a corresponding
  mixing-length model with $\mlp=1.5$ (grey solid line, red in color version).
  \ion{Ti}{i} lines are also see to absorb in the 8432 epsilon subband at
  $\lambda\lambda$\,8435.7,\,8435.0\pun{\AA}, and bluewards of the triple band
  system at 8412.3 and 8426.5\pun{\AA}.  Another strong \ion{Ti}{i} line
  is also absorbing at 8468.4\pun{\AA} (not seen in this plot).}
\label{f:TiO}
\end{figure}

\begin{figure}
\resizebox{\hsize}{!}{\includegraphics[draft = \draftflag]%
{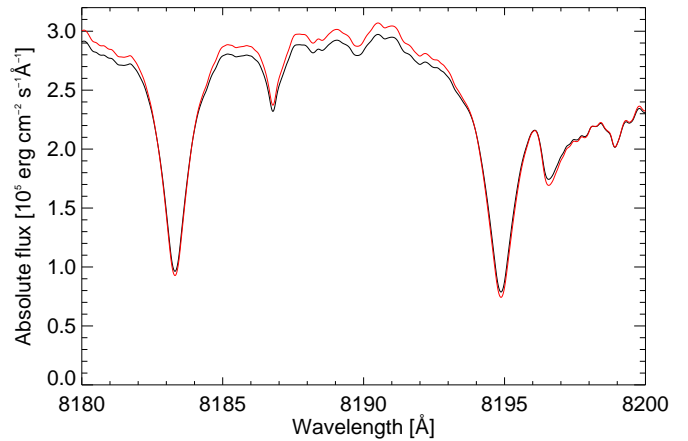}}
\caption[]{Comparison of synthetic spectra showing a subordinate \ion{Na}{i}
  doublet (at 8183.3,
  8194.8\pun{\AA}) based on hydrodynamical model~\mocl\ (black solid line) and
  a corresponding mixing-length model with $\mlp=1.5$ (grey solid line, red in
  color version).}
\label{f:NaI}
\end{figure}
\end{samepage}

We find that, accounting for the hydrodynamical structure, yields
systematically weaker TiO (and H$_2$O not shown) bands by 0.18 and 0.025\,dex
respectively, while the pseudo-continuum appears unchanged.  This is possible
because the strongest TiO bands are formed at two dex lower optical depth than
the opacity minima between those bands.  The differences in the strength of
the epsilon subband heads in the synthetic spectra of Fig.~\ref{f:TiO} would
correspond to a difference in \Teff\ of $\approx 200\pun{K}$ when compared to
an observed spectrum, in the sense that an analysis based on MLT models
overestimates \Teff.

Atomic lines absorbing through TiO bands troughs such has the doublets of Ti
at $\lambda\lambda$\,8435.7,\,8435.0\pun{\AA}, of \ion{K}{i} at
$\lambda\lambda$\,7664.9,\,7699.0\pun{\AA}, and all other atomic lines formed
bluewards of 0.7 $\mu$m, would look wider and deeper in contrast to the TiO
pseudo-continuum, causing MLT models to overestimate gravities.  This is not
the case of the \ion{Na}{i} doublet at
$\lambda\lambda$\,8183.3,\,8194.8\pun{\AA} shown in Fig.~\ref{f:NaI} which is
practically unaffected by this pseudo-continuum because it forms between TiO
and VO band heads from deeper photospheric layers.  The same would be true of
lines formed around the peak of the spectral distribution between 0.9 and
1.3\pun{$\mu$m}.  Although of course these can be as well affected in analysis
where \Teff\ is determined from the TiO bands.

} 


\section{Mixing by atmospheric overshoot}
\label{s:mixing}

For describing the mixing properties of the flow field in the overshooting
layers of our models, we follow the approximate procedure laid out
by LAH. We describe the mixing in terms of a mass exchange frequency given 
by
\beq
\fex(z) \equiv \frac{\langle\Fmup\rangle(z)}{\langle\mcol\rangle(z)}.
\label{e:fex}
\eeq
\changed{\Fmup\ is the upward directed component of the mass flux 
\beq
\Fmup(x,y,z,t)\equiv\left\{
\begin{array}{ll}
\rho v_\mathrm{z} & \mathrm{if}\blank v_\mathrm{z} > 0\\
0                 & \mathrm{otherwise}.
\end{array}\right.
\eeq
where $v_\mathrm{z}$ is the vertical component of the velocity (counted as
positive if directed upwards), $\rho$ the mass density, $x$, $y$, $z$ the
spatial coordinates, and $t$ the time. 
\mcol\ is the mass column density given by
\beq
\mcol(x,y,z,t)\equiv\int_{z}^{\infty}\! dz^\prime\,\rho(x,y,z^\prime,t).
\eeq
\mbox{$\langle.\rangle$} denotes the horizontal and temporal average over $x$,
$y$, and $t$. }
The basic idea is to take the time scale over which the mass above a certain
reference height is potentially exchanged by the flow as time scale over
which material is mixed with fresh material stemming from the deeper lying,
convective layers.  As we shall see the mass exchange frequency \fex\ exhibits
an exponential height dependence. The mixing rate given by
relation~\eref{e:fex} is an approximation only. Depending on the way the
mixing takes place in detail, the normalization of the mixing profile might
change. However, the relative shape of the mixing rate -- the exponential
decline -- is a robust feature, and in the following we shall characterize the
mixing found in our models by the scale height of the exponential decline.

We note that in \citet{Ludwig03} the mixing was described in terms of a mixing
velocity
\begin{equation}
\vmix \equiv \frac{\langle\Fmup\rangle}{\langle\rho\rangle}
\end{equation}
where $\rho$ denotes the mass density. \vmix\ also shows an exponential height
dependence, and its rate of decline was given in the above paper. While at the
level of our approximation the description in terms of \vmix\ is equivalent to
the one in terms of \fex, the scale heights of the various declines can differ
substantially. Differences become large in cases where the scale heights of
\vmix\ or \fex\ are large, or the scale heights of $\rho$ and \mcol\ differ
noticeably.

\subsection{Subsonic filtering}

The atmospheric velocity field is a superposition of advective
motions and acoustic waves generated by convection in deeper layers \citep[see
also LAH, and ][]{Ludwig+Nordlund00}. The wave motions contribute little if at
all to the mixing due to their spatially coherent, oscillatory character. The
overshooting, convective motions tend to decay with distance to the
Schwarzschild stability boundary, while the wave amplitudes tend to increase
with height due to the sharp decrease in mass density. This leads to the
situation that beyond a certain height the atmospheric velocity field becomes
dominated by wave motions. In order to get a reliable estimate of the mixing,
it is therefore necessary to remove the wave contributions to the velocity
field before evaluating the mass flux~\Fmup.

\begin{figure}[!tb]
\resizebox{\hsize}{!}{\includegraphics[draft = \draftflag]%
{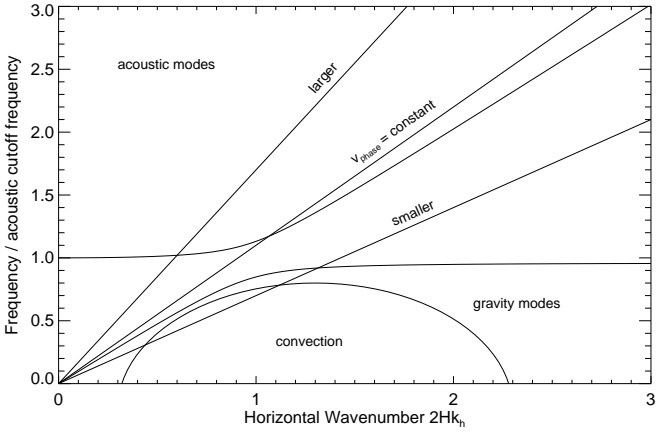}}
\caption[]{Schematic illustration of subsonic filtering in the $k$-$\omega$ 
domain: only components
in wavenumber-frequency domain below a prescribed phase
speed~$v_{\mathrm{phase}}$ are retained. They preferentially belong to
convective motions.}
\label{f:subsonic}
\end{figure}

We removed the wave contributions by subsonic filtering -- a technique
developed in the context of solar observations for cleaning images from
``noise'' stemming from the solar 5\,minute oscillations \citep{Title+al89}.
Figure~\ref{f:subsonic} schematically illustrates this filtering technique.
In short, one considers a time sequence of images and removes features with
horizontal phase speeds~$v_{\mathrm{phase}}$ greater than a prescribed
threshold. This is achieved by Fourier filtering of spatial-temporal data in
the $k$-$\omega$ domain.  For every depth layer in our data cubes we performed
a 3D Fourier analysis (one temporal, two spatial dimensions) of the vertical
mass flux retaining only contributions below a preset phase speed threshold.
In practice, acoustic and convective contributions are not as cleanly
separated as shown in the Fig.~\ref{f:subsonic}, and one must find the right
balance between removing as much acoustic components as possible while
retaining as much as possible convective contributions. We always studied a
sequence of phase speed thresholds in order to judge the success of the
procedure.

\subsection{Mixing in the solar atmosphere}

\begin{figure}[!tb]
\resizebox{\hsize}{!}{\includegraphics[draft = \draftflag]%
{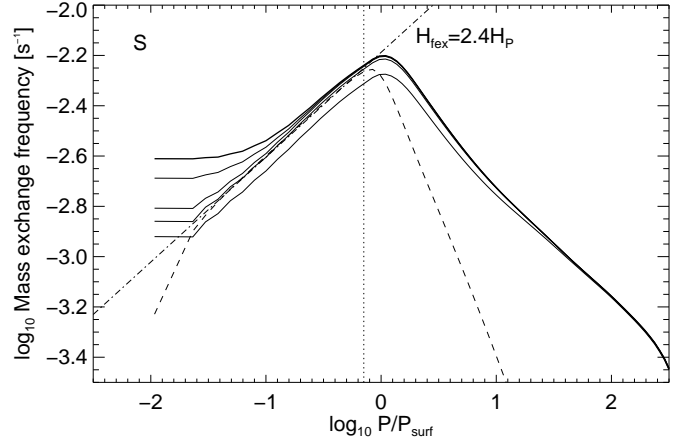}}
\caption[]{Mass exchange frequency~\fex\ in solar model \mosu\ as a function of gas
  pressure: the unfiltered data (thick solid line) were subsonically filtered
  retaining only features with phase speeds $v_{\mathrm{phase}} <$ 12, 6, 3,
  and 1.5\punkms\ (thin solid lines from top to bottom). The approximate
  location of the Schwarzschild boundary of convective stability is indicated
  by the dotted vertical line, \fex\ of a convective eigenmodes with
  horizontal wavelength $\lmode=5.0\pun{Mm}=33\,\Hpsurf$ by the dashed line.
  The dash-dotted line is a fit depicting the decline of \fex. It is labeled
  by the scale height of the decline in units of the local pressure scale
  height. The pressure is given in units of the pressure at Rosseland optical
  depth unity~\Psurf. The plateau at lowest pressures is an artifact of the
  upper boundary condition applied in the RHD model.}
\label{f:Spmix}
\end{figure}

For reference we begin with a discussion of the mixing in the solar
atmosphere.  Figure~\ref{f:Spmix} shows \fex\ in our solar model for various
degrees of subsonic filtering. It is clearly visible that the subsonic
filtering has the strongest impact on \fex\ in the uppermost atmospheric
layers. As hinted above, {\em \fex\ exhibits an exponential decline
  with height (\/$\log P \propto z$) after appropriate subsonic filtering\/}
which was put forward by \citet{Freytag+al96} as generic feature of convective
overshoot.  Figure~\ref{f:Spmix} further shows that too low a velocity
threshold removes also convective features, as visible by the reduction of the
velocity in the deeper, convection dominated layers for the case of a phase
speed threshold of 1.5\punkms.  Quantitatively, for the Sun we find a
scale height of \fex\ in terms of the local pressure scale height of 
$\Hfex=2.4\,\Hp$ with an uncertainty\footnote{The uncertainty is not meant in
  a statistical sense but reflects the precision with which we can read off
  the slope from the plots of \fex. The alert reader might suspect a
  connection to the ``chi-by-eye'' technique, consult \citet{Press+al92} for a
  discussion of the immediate consequences.} of about 10\,\%.

Apart from purely numerical findings, an exponential decline of \fex\ is also
motivated from semi-analytical considerations of the behavior of linear
convective modes \citep{Freytag+al96}.  In Fig.~\ref{f:Spmix} we plotted the
\fex-profile of a linear convective eigenmode with horizontal wavelength of
$\lmode=5.0\pun{Mm}$. We used the temporally and horizontally averaged
hydrodynamical structure as background on which we solved the linearized
hydrodynamical equations.  The absolute amplitude of the mode has been scaled
to match \fex\ of the hydrodynamical model leaving the shape of the mode's
\fex-profile intact.  The exchange frequency of the mode exhibits an
exponential ``leakage'' into the formally convective stable layers. Generally,
the rate of decline depends on \lmode, being faster for modes of shorter
horizontal wavelengths. The mode with 5.0\pun{Mm} wavelength was chosen since
it provided a good overall fit to the decline of \fex\ in the hydrodynamical
model. The wavelength of this mode is significantly larger than the horizontal
scale of the dominant convective structures on the Sun -- the granules with
typical sizes of around 1.2\,Mm.  This might be related to the assumption of
adiabaticity in the mode calculations, which is not a good approximation in the
solar photosphere, or to the fact that a convective mode is a non-stationary
solution of the hydrodynamical equations.

From the rather large wavelength of the best fitting mode one might argue that
5.0\pun{Mm} is close to the geometrical size of the computational box of
model~\mosu\ (6.0\pun{Mm}), and actually the box size sets the rate of decline
of \fex. We verified that a solar model of about twice the horizontal size
gives the same rate of decline as model~\mosu. The box size of model~\mosu\ is
sufficient to allow the build-up of all convective structures contributing
significantly to the overshooting velocity field in the deep photosphere.  The
box sizes of the M-type models are allowing the presence of a similar
number of convective cells as the solar model. Thus, we expect that also our
M-type models capture the relevant convective structures controlling the
overshooting motions.

\subsection{Mixing in M-type atmospheres}

\begin{figure}
\resizebox{\hsize}{!}{\includegraphics[draft = \draftflag]%
{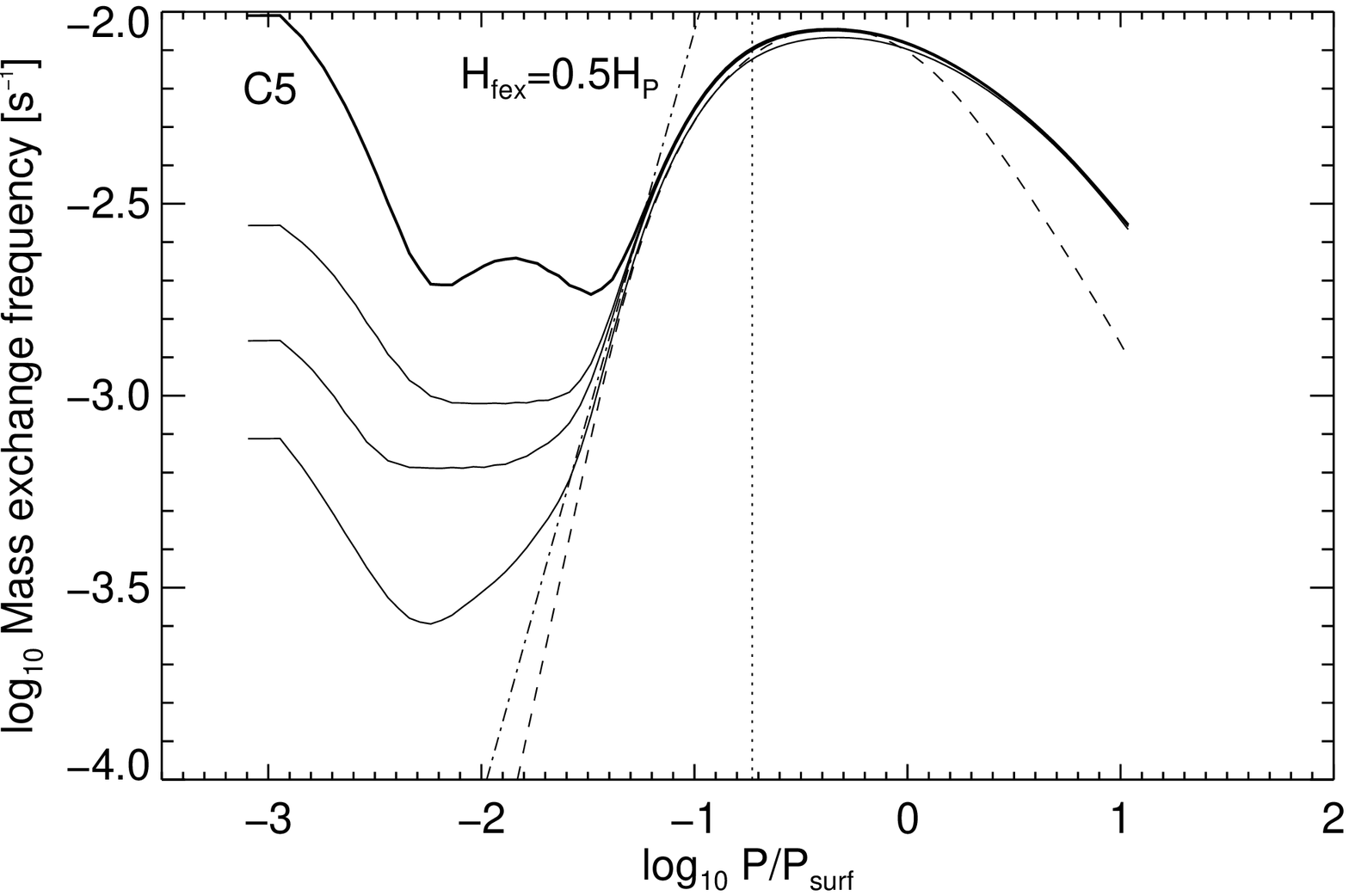}}
\caption[]{Like Fig.~\ref{f:Spmix}, model~\moch.
  $v_{\mathrm{phase}} <$ 1.0, 0.5, 0.25\punkms, $\lmode=250\pun{km}=22\,\Hpsurf$.
\label{f:C5pmix}}
\end{figure}

\begin{figure}
\resizebox{\hsize}{!}{\includegraphics[draft = \draftflag]%
{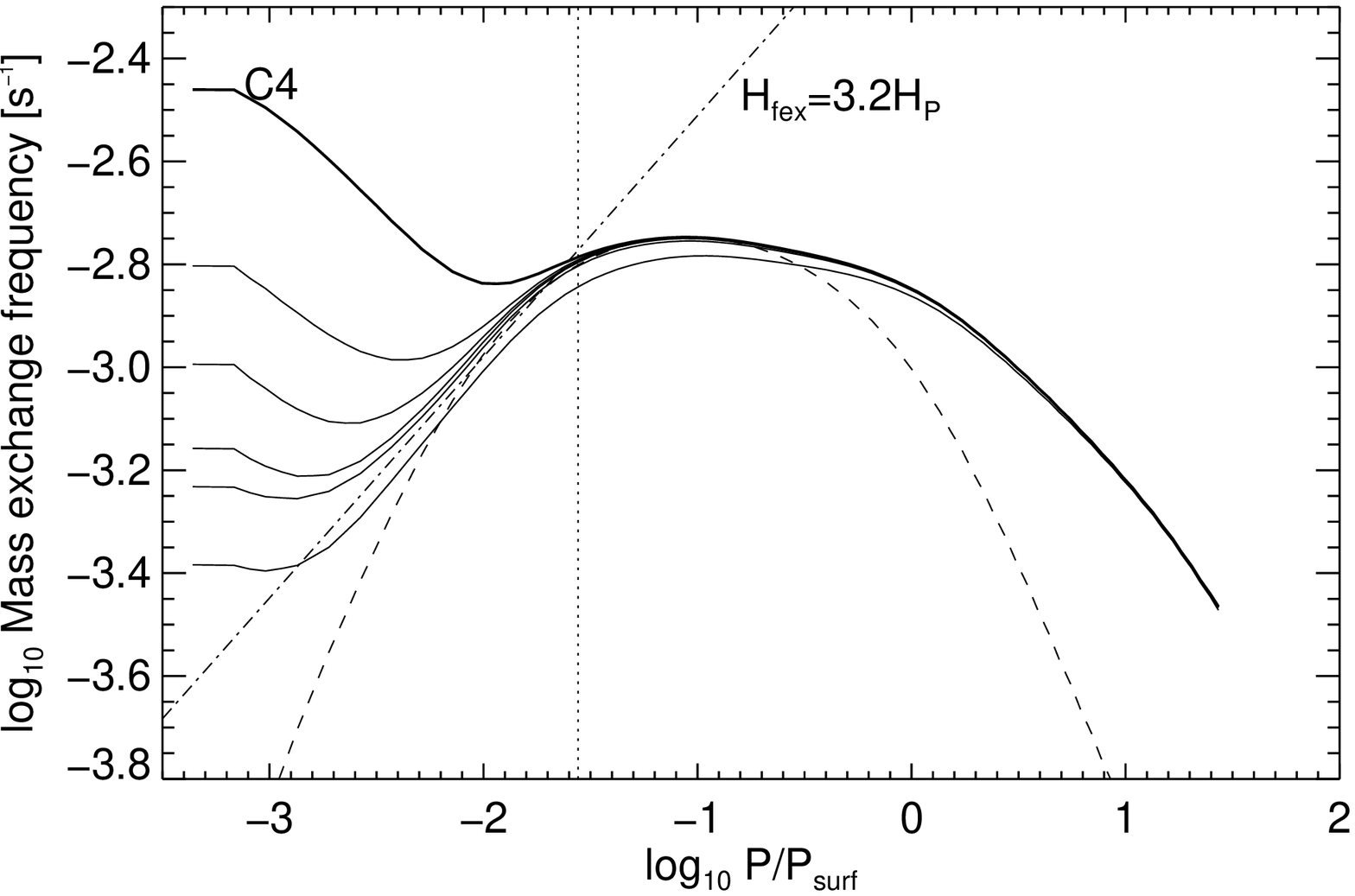}}
\caption[]{Like Fig.~\ref{f:Spmix}, model~\mocm.
  $v_{\mathrm{phase}} <$ 8.0, 4.0, 2.0, 1.0, 0.5\punkms, $\lmode=3.0\pun{Mm}=24\,\Hpsurf$.
\label{f:C4pmix}}
\end{figure}

In Fig.~\ref{f:C5pmix} and~\ref{f:C4pmix}, we show the vertical distribution of the
mass-exchange rates \fex\ for two of our M-type models.  With decreasing \logg\ 
and increasing \Teff, the zone of convective instability extends further and further 
into the optically thin atmosphere, leaving little room for overshoot in the models~\mocl\ 
and~\mohm.  Reading off an exponential decline rate is very uncertain in these
models. However, from them we find a slow decline with $\Hfex\approx 18\,\Hp$
in \mocl\ and $\approx 28\,\Hp$ in \mohm.  Models~\mocl\ and~\mocm\ leave more
room for overshooting, allowing a more precise determination of the exponential 
decline. We estimate the involved uncertainty to about 20\,\%. We find $\Hfex=0.5\,\Hp$ 
in model ~\moch\ and $3.2\,\Hp$ in~\mocm.  Note, that in the model~\moch\ with 
highest gravity and steepest decline of \fex\ the exponential behavior does not 
set in immediately at the boundary of convective stability

Qualitatively, in terms of the rate of decline, overshooting is less pronounced 
in models of higher gravity. This is in part due to the fact that buoyancy forces 
scale proportional to gravity, making buoyancy more effective in confining the 
convective motions to the formally unstable regions. At lower gravity, mixing -- 
despite the increasing geometrical scales -- is more rapid, not only due to the 
slower decline of the mixing rate but also due to the higher convective velocities.

Silicate cloud formation is one of the most important aspect of the modeling
of late-type M and brown dwarfs.  The formation of clouds is understood as a
compromise between condensation, sedimentation and advection (turbulent
overshooting mixing) time scales which determine the extension, location in
the thermal atmospheric structure, and composition of the cloud deck.  To
represent the correct distribution with height of the mixing time scale,
investigators have experimented with various descriptions for \fex:
\citet{Allard+al03} used a parabolic function with opening set by the
innermost and outermost convective layers, and normalized at the convective
velocity maximum, while \citet{Ackerman+Marley01} and \citet{Cooper+al03}
preferred a constant distribution throughout the atmosphere, set to the value
associated with the maximum of the convective velocity.  A modeling with
convective modes should give a more physical description which could be
implemented in 1D model atmospheres. Trying to match the mixing profiles in
the overshooting regions with convective modes, however, worked only partially
so far.  For the lower gravity models the fits were not satisfactory.  This
might be related to the situation that convection reaches high up, and we do
not actually see the asymptotic exponential tail of the mixing profile.  We
oriented the horizontal wavelength of the linear modes at the largest sizes of
structures the computational box could accommodate in the respective models.
Despite the present shortcoming we are optimistic that one can add refinements
to the mode-modelization that would allow to satisfactorily match the RHD
results.

If the mixing trends observed in our models hold for cooler objects, these
go in the direction of making clouds thicker or more extended into higher
atmospheric layers with decreasing gravities and increasing \Teff\ .  However,
decreasing pressure will work in the opposite direction, making it harder for
grains to form. Detailed calculations will be presented in a subsequent
publication.  Nevertheless, we expect that clouds will be more extended for
young objects than for older ones of same \Teff, and that these will remain
dusty at lower~\Teff\ and later spectral types, i.e, below spectral class T4 or 1400\,K
\citep{Golimowski+al04}.

\subsection{Cloud cover disruption in early T-type brown dwarfs}

\citet{Burgasser+al03} have found a resurgence of molecular spectral features
such as FeH bands in the spectra of early T-type brown dwarfs. This is
interpreted as a spectral signature of the onset of cloud cover disruption.
Indeed, these spectral features of refractory species can only be seen if the
atmosphere is transparent enough to observe flux emerging from below the cloud
forming layers. This is possible if holes in the cloud deck are occurring.

Dust does not form in the models studied in this work. However, here we want
to speculate how the cloud pattern might look when one expects a disrupted
cloud layer like in early T-dwarfs.  The cloud deck is shaped by convective
overshooting which mixes up refractory material into the grain condensing part
of the atmosphere -- below gas temperatures of 2000\pun{K}, in early T-dwarfs
perhaps over one pressure scale height above the convectively unstable layers.
This far above the convection zone, the horizontal and vertical motions are
not correlated in the same way as in the strongly convective layer where the
flow forms cell-like patterns.  Structures larger than the granular scale in
combination with waves dominate the velocity field. The typical granular flow
pattern is "washed out" from the flow higher up. Hence, we do not expect that
the cloud deck is fragmented on a spatial scale given by the granular scale,
but likely on a larger scale. This consideration refers to effects of
convection.  It is of course well possible that the actual cloud pattern is
rather shaped by the global wind circulation expected to be present in
rotating brown dwarfs or planets.

\section{Final remarks}
\label{s:remarks}

We have seen that mixing-length theory provides a reasonably realistic picture
of the convective energy transport in M-type atmospheres, even considering
that a substantial part of the optically thin atmosphere is affected by
convection.  However quantitatively, temperature errors of up to $\approx
250\pun{K}$ \changed{(or 9\,\%)} are possible if one (unluckily) picked a
value of unity for the mixing-length parameter (\mlp) entering MLT.  The
efficiency of the convective energy transport measured in terms of an
effective mixing-length parameter is rather high in M-type atmospheres.
Choosing a larger \mlp\ helps but is not sufficient to describe the thermal
structure of M-type atmospheres if one wishes to attain a high level of
accuracy. MLT does not provide the precise scaling of the convective transport
efficiency with stellar parameters and optical depth.  To get a better
quantitative description one might try to calibrate besides \mlp\ the
``internal'' parameters of MLT with radiation-hydrodynamics models. M-type
atmospheres appear particularly well suited for this undertaking since
convection takes place under optically thick and thin conditions.

Our results have shown that convective overshooting mixes the layers of M-type
atmospheres which are formally (according to the Schwarzschild criterion)
stable against convection much more strongly towards higher effective
temperatures and lower gravities.  This should make young late-type M-dwarfs
and brown dwarfs even more cloudy than older disk objects of the same \Teff.
We think that a detailed modeling of such cooler atmospheres, especially
around 1400\pun{K} should deliver important clues about the interesting
question of the cloud cover, and should help to understand the break in colors and
spectral type vs. effective temperature relation observed for these objects
\citep{Knapp+al04,Golimowski+al04}.

To aid the modeling of the spectral properties of cloudy objects, we further
think it should be worthwhile to improve the model of convective modes which
where mostly used for demonstration purposes in this work. Our current mode
model did not perform sufficiently to be up to the task but a number of
refinements can be brought to this model and tested. It could be calibrated
with RHD models in a similar fashion like MLT. At a higher ambition level one
might even contemplate to combine ideas into a single description for the
energy transport and effects related to overshooting. We are aware that many
attempts have been made to improve or even completely replace MLT since it has
been introduced into astrophysics in the early 1950s by B\"ohm-Vitense -- with
mixed success. Our goal would not be to formulate a new convection theory but
rather a parameterized model like MLT which is flexible enough to fully fit
RHD results, and contains sufficient physics to allow a robust inter- and
extrapolation in a wide range of stellar parameters. The availability of
detailed RHD models appear essential to identify the necessary building
blocks.

The main uncertainty affecting our present results is related to the
approximate treatment of the wavelength-dependence of opacity in the radiative
transfer which was optimized for an atmosphere at \Teff=2900\,K and \logg=5.3.
So, one of the first issues to be addressed in future work is the improvement
of this approach.  Work is under way for a refined implementation in a new 3D
radiation-hydrodynamics code
\citep[named \mbox{CO$^5$BOLD},][]{Freytag+al02,Wedemeyer+al04}. We finally
emphasize that our results apply to atmospheres of solar metalicity. We
expect marked differences for metal-poor atmospheres \citep[see, e.g.,][]{Asplund+al99b}.

\begin{acknowledgements}
  The authors are indebted to Isabelle Baraffe and Gilles Chabrier for their
  supportive enthusiasm during the course of the project, and their scientific
  input during numerous discussions.  HGL would like to thank {\AA}ke Nordlund
  and Robert Stein for making available a version of their hydrodynamical
  atmosphere code, as well as Frank Robinson for providing unpublished data of
  his convection models. HGL further acknowledges financial support of the
  Walter Gyllenberg Foundation in Lund and the Swedish Research Council. PHH
  was supported in part by the P\^ole Scientifique de
  Mod\'elisation Num\'erique at the ENS-Lyon. Some of the calculations presented
  here were performed at the H\"ochstleistungs Rechenzentrum Nord (HLRN), and
  at the National Energy Research Supercomputer Center (NERSC), supported by
  the U.S. DOE.  We thank all these institutions for a generous allocation of
  computer time.
\end{acknowledgements}

\appendix

\changed{%
\section{Computation of the P{\'e}clet number}
\label{s:peclet}

The P{\'eclet} number~\Pe\ measures the relative importance between conductive
(here by radiation) and advective heat transport  
\beq
\Pe \equiv \frac{t_\mathrm{rad}}{t_\mathrm{adv}}. 
\label{e:peclet}
\eeq
$t_\mathrm{rad}$ is a radiative relaxation time, and $t_\mathrm{adv}$ a
characteristic time over which the temperature of moving gas elements
changes due to adiabatic compression or expansion. In the present context, we
employ a mixing-length picture and evaluate the radiative relaxation
time~$t_\mathrm{rad}$ with the MLT formula
\beq
t_\mathrm{rad} = \frac{\rho\cp\lmix\taueddy}{f_3\sigma\chi T^3} \,
                    \left( 1+\frac{f_4}{\taueddy^2}\right).
\label{e:tauradi}
\eeq
\cp\ denotes the specific heat at constant pressure, $\lmix=\mlp\Hp$
the mixing-length, $\sigma$ Stefan-Boltzmann's constant, $\chi$ opacity, $T$
temperature, $\rho$ mass density, and $\taueddy$ the optical thickness of a
convective element defined as
\beq
\taueddy\equiv\chi\rho\lmix.
\eeq
$f_3=16$ and $f_4=2$ are dimensionless constants set to values assumed in the
MLT formulation of Mihalas \citep[see][]{Ludwig+al99}. We further
assume a mixing-length parameter~$\mlp=2.5$ which is a reasonable value for
the M-type atmospheres under consideration (see Fig.~\ref{f:entropy}).

Similarly, we estimate $t_\mathrm{adv}$ as the time interval over which a vertically moving
gas element has build up a substantial temperature difference according to
\beq
t_\mathrm{adv} = \frac{\lmix}{v_\mathrm{c}}
\eeq
where $\nabla$ is the logarithmic temperature derivative of the thermal
profile with respect to pressure, $\nabla_\mathrm{ad}$ the corresponding
adiabatic value, and $v_\mathrm{c}$ a convective velocity we set to a typical
atmospheric value (300\pun{m s$^{-1}$} for model \moch\ and 600\pun{m
  s$^{-1}$} for model~\mocl, see Fig.~\ref{f:velmlt}). Note, that in this
paper we argue taking recourse to \textit{ratios} of \Pe\ only. This
makes the precise choice of arbitrary or little constrained parameters
less critical. 
} 

\bibliographystyle{aa}
\bibliography{4010.bib}

\end{document}